\documentclass[showpacs,preprintnumbers,amsmath,amssymb,superscriptaddress,nofootinbib]{revtex4}
\usepackage{graphicx}
\usepackage{dcolumn}
\usepackage{bm}
\usepackage{longtable}
\usepackage[countmax]{subfloat}
\usepackage{epstopdf}
\usepackage{easybmat}

\providecommand{\be}{\begin{equation}}
  \providecommand{\ee}{\end{equation}}
\providecommand{\bea}{\begin{eqnarray}}
  \providecommand{\eea}{\end{eqnarray}}
\providecommand{\ba}{\begin{eqnarray}}
  \providecommand{\ea}{\end{eqnarray}}
\providecommand{\beas}{\begin{eqnarray*}}
  \providecommand{\eeas}{\end{eqnarray*}}

\providecommand{\beni}{\begin{equation*}}
  \providecommand{\eeni}{\end{equation*}}

\providecommand{\bw}{\begin{widetext}}
  \providecommand{\ew}{\end{widetext}}

\usepackage{epsfig}
\usepackage{amssymb}
\usepackage{amsmath}
\usepackage{amsthm}
\usepackage{latexsym}
\def\be{\begin{equation}}
\def\ee{\end{equation}}
\def\bc{\begin{center}}
\def\ec{\end{center}}

\def\be{\begin{equation}}
\def\ee{\end{equation}}
\def\bea{\begin{eqnarray}}
\def\eea{\end{eqnarray}}

\begin{document}

\preprint{APS/123-QED}

\title{First-passage phenomena in hierarchical networks}

\author{Flavia Tavani}
\affiliation{Dipartimento SBAI (Ingegneria), Sapienza Universit\`a di Roma, via A. Scarpa 16, 00161, Roma, Italy.}
\date{\today}

\author{Elena Agliari}
\affiliation{Dipartimento di Matematica, Sapienza Universit\`a di Roma, P.le A. Moro 2, 00185, Roma, Italy.}

\begin{abstract}
In this paper we study Markov processes and related first passage problems on a class of weighted, modular graphs which generalize the Dyson hierarchical model. In these networks, the coupling strength between two nodes depends on their distance and is modulated by a parameter $\sigma$. We find that, in the thermodynamic limit, ergodicity is lost and the ``distant'' nodes can not be reached. Moreover, for finite-sized systems, there exists a threshold value for $\sigma$ such that, when $\sigma$ is relatively large, the inhomogeneity of the coupling pattern prevails and ``distant'' nodes are hardly reached. The same analysis is carried on also for generic hierarchical graphs, where interactions are meant to involve $p$-plets ($p>2$) of nodes, finding that ergodicity is still broken in the thermodynamic limit, but no threshold value for $\sigma$ is evidenced, ultimately due to a slow growth of the network diameter with the size. 
\end{abstract}
 \maketitle

\section{Introduction}

In most real-life networks links are associated to weights accounting for  wiring costs that typically depend on the distance between nodes to be connected. 
According to the system considered, such a distance may be a physical distance, a social distance, or any quantity which measures the cost associated with the formation of a link \cite{Barthelemy-PhysRep2011}. For instance, in the brain, since axons are expensive in terms of material and energy, regions that are spatially closer have a greater probability of being connected than remote regions \cite{Bullmore-Nature2009, Friston-Science2009}. Also, in lymphocyte networks, clones that display a larger affinity (i.e, that are closer in the idiotypic phase space) are more likely to experience a mutual regulation \cite{BA-JStat2010, Agliari-EPL2011}.

Here, we consider a network embedded in an ultrametric space, where the coupling $J_{ij}$ between two nodes $i$ and $j$ scales as $J_{ij} \sim 2^{-2 \sigma d_{ij}}$, being $d_{ij}$ the distance between $i$ and $j$, and $\sigma$ a proper, positive tunable parameter. This network was originally introduced by Dyson in the 1960s to describe non-mean-field spin systems \cite{Dyson-CMP1969} and, more recently,  also the Sherrington-Kirkpatrick model for spin glasses \cite{Castellana-JSP2014,Castellana-PRL2010,Castellana-PRE2011} and the Hopfield model for neural networks \cite{ABGGTT-JPA2015,ABGGTT-NN2015} defined on such a topology have been investigated. Indeed, the peculiar features (e.g., high degree of modularity) of such a weighted graph play a crucial role in the statistical-mechanics treatability as well as in the emergent behavior of the above mentioned models \cite{ABGGTT-PRE2015}. 

Here, we proper generalize the original model in order to account for systems exhibiting $p$-wise interactions (e.g., $p$-spin systems). Basically, as we are going to explain, this implies a new parameter $p$ tuning the size of the modules making up the system and, accordingly, $J_{ij} \sim p^{-d_{ij}[p-2(1-\sigma)]}$.

Our goal is to investigate and to quantify the influence of the pattern of weights on
dynamical processes occurring on the network. In particular, we focus on Markovian processes where the probability to move from a state $i$ to a state $j$ is given by $J_{ij}$ (upon proper normalization) and we calculate the first passage probability and the mean first passage time \cite{Redner-2001Book} to reach a given node or a given set of nodes as a function of its distance from the starting point.

We find that in the thermodynamic limit, ergodicity is broken for any value of $p$, meaning that in the limit of infinite size distant nodes can not be reached, no matter how long the process is run.
Moreover, by comparing the probability that the process jumps on ``close'' nodes with the probability that the process jumps on ``distant'' nodes, we find that, as long as $p=2$ (i.e. the basic modules are made by two nodes), there exists a threshold value for $\sigma$, such that when $\sigma$ is relatively large the inhomogeneity induced by distance-depending weights is strong and 
the process actually tends to stay in the nighbourhood of the starting node, while when $\sigma$ is relatively small the inhomogeneity is less effective and the process can move away from the initial module.
On the other hand, when $p>2$ (i.e., the basic modules are made by cliques of size $p$), inhomogeneity is always effective for the process which is always (regardless of $\sigma$ and of the size) more likely to wonder in the neighborhood of the starting module.
A qualitative difference between the case $p=2$ and the case $p>2$ is highlighted also while looking at the mean time to first reach the farthest node and the closest node, respectively. In fact, when $p=2$ the ratio between these quantities remains finite for any $\sigma$ but $\sigma =1$ (in this case the ratio grows logarithmically with the system size), while when $p>2$ the ratio is always increasing with the system size.

The paper is organized as follows. In Sec.~\ref{sec:defintions} we review the growing algorithm for the network under study and we define the transition matrix of the related Markov process. Sections \ref{sec:asymptotics2}-\ref{sec:hitting2} are devoted to the case $p=2$: we first analyze the asymptotic properties of the stochastic process, then we move to the estimate of mean first passage times and of the splitting probabilities. Analogous calculations are performed for the general case $p>2$ in Secs.~\ref{sec:asymptoticsp} and \ref{sec:hittingp}. Finally, Sec.~\ref{sec:conclusions} is left to discussions and outlooks. The most technical details and lengthy calculations are collected in the Appendices. 

\section{Definition of the models} \label{sec:defintions}

In this work we focus on deterministic, weighted, recursively grown graphs, referred to as $\mathcal{G}^{(p)}$, originally introduced to embed statistical-mechanics spin systems. We shall consider not only graphs corresponding to purely pairwise  ($p=2$) interactions among spins \cite{Dyson-CMP1969}, but we are generalizing the structure in order to account for $p$-wise ($\mathbb{N} \ni p\geq 2$) interactions as well \cite{Castellana-JSP2014b}. 

Such systems can be formalized through Hamiltonians defined recursively, in such a way that at the first iteration one has a set $\{S\}_1=\{S_{i_1},S_{i_2},...,S_{i_p}\}$ of $p$ spins coupled together; at the second iteration one takes $p$ replicas of the previous system and couples the related spins hence obtaining a system containing $p^2$ spins referred to as $\{S\}_2$, and so on for the following iterations. In particular, for the ferromagnetic case one has
%
\begin{eqnarray}\label{derrida}
H_0(\{S\}_0 |\sigma) &=& 0, \\
\label{eq:H2}
H_1(\{S\}_1 |\sigma) &=& -\frac{1}{p^{[p-2(1-\sigma)]}}\sum_{i_1<\cdots <i_p}^{p}S_{i_1}\cdots S_{i_p}, \\ 
H_2(\{S\}_2 |\sigma) &=& \sum_{l=1}^{p}H_{1}(\{S_l\}_1|\sigma)-\frac{1}{p^{2[p-2(1-\sigma)]}}\sum_{i_1<\cdots <i_p}^{p^2}S_{i_1}\cdots S_{i_p}, \\
\nonumber
\vdots & & \vdots \\
\label{eq:Hp}
H_K(\{S\}_K |\sigma) &=& \sum_{l=1}^{p}H_{K-1}(\{S_l\}_{K-1}|\sigma)-\frac{1}{p^{K[p-2(1-\sigma)]}}\sum_{i_1<\cdots <i_p}^{p^K}S_{i_1}\cdots S_{i_p}.
\end{eqnarray}
%
Spins are binary and take values $+1$ or $-1$. The parameter $\sigma$ is bounded as $\sigma \in (1/2, \ 1]$: for $\sigma >1$ the interaction energy goes to zero in the thermodynamic limit, while for $\sigma \leq 1/2$ the interaction energy diverges in the same limit. Also, notice that the coupling among spins is positive due to the ferromagnetic nature of the model which makes neighboring spins to ``imitate'' each other.

\begin{figure}[tb]
\includegraphics[width=10cm]{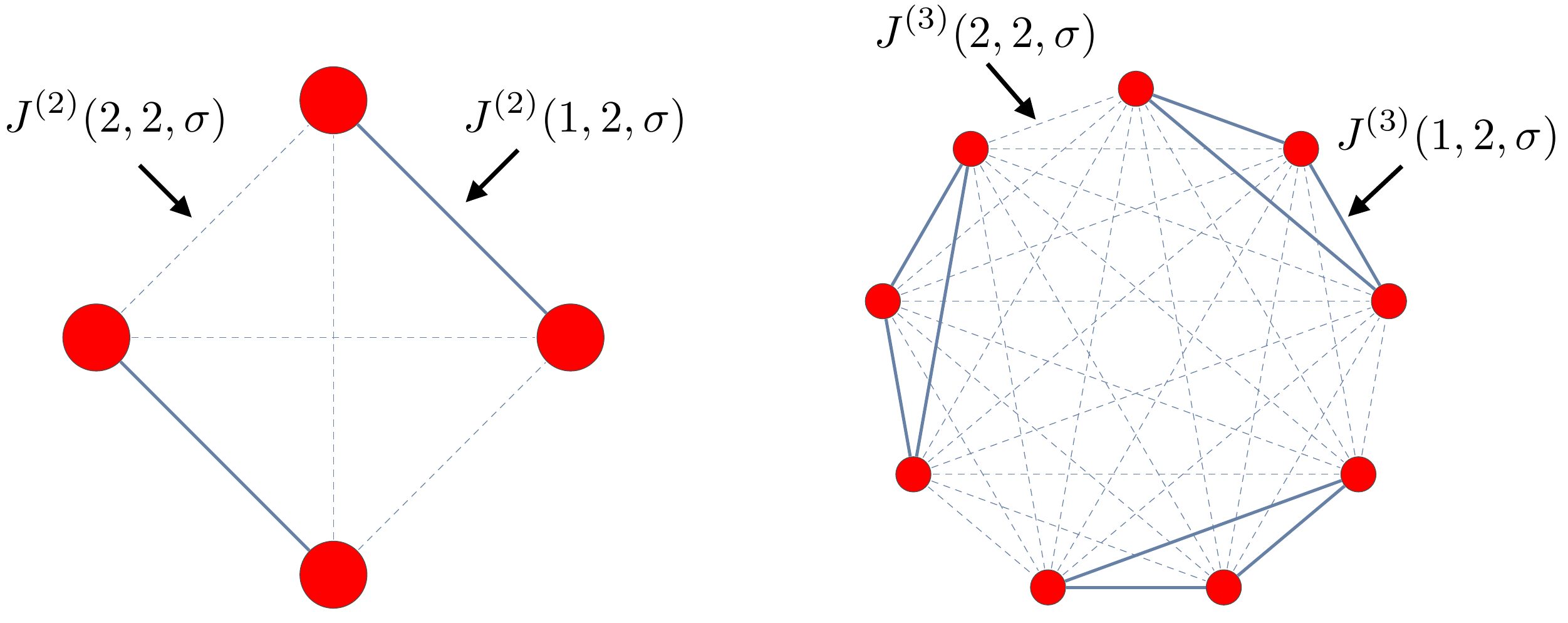}
		\caption{(Color
			online) Example of $\mathcal{G}^{(p)}$, with $p=2$ (left) or $p=3$ (right) and $K=2$. The solid and thicker lines represent links with a larger weight, while the dashed ones are the links added at the second iteration and therefore carrying a lower weight. }\label{fig:G}
\end{figure}

Now, the graph underlying such a system can as well be built iteratively (see \cite{ABGGTT-PRE2015} for the special case $p=2$). The construction begins with $p$ nodes, fully connected with links carrying a weight $J^{(p)}(1,1,\sigma) = p^{-[p -2(1-\sigma)]}$. We refer to this graph as $\mathcal{G}_1^{(p)}$. At the next step, one introduces $p$ replicas of $\mathcal{G}_1^{(p)}$ and connects the nodes pertaining to different replicas with links displaying a weight $J^{(p)}(2,2,\sigma) = p^{-2[p +2(1-\sigma)]}$; also, the weight on the existing links is updated as $J^{(p)}(1,1,\sigma) \rightarrow J^{(p)}(1,2,\sigma) = J^{(p)}(1,1,\sigma) + J^{(p)}(2,2,\sigma)$. The graph $\mathcal{G}^{(p)}_2$ counts now $p^2$ nodes.  At the generic $k$-th iteration, one introduces $p$ replicas of $\mathcal{G}_{k-1}^{(p)}$, insert $p^{2k-1}$ new links, each carrying a weight $J(k,k,\sigma)=p^{-k[p +2(1-\sigma)]}$, among nodes pertaining to different replicas, and the weights on existing links are updated as $J^{(p)}(d,k-1,\sigma) \rightarrow J^{(p)}(d,k,\sigma) = J^{(p)}(d,k-1,\sigma) + J^{(p)}(k,k,\sigma)$, for any $d<k$. If we stop the iterative procedure at the $K$-th iteration, the final graph $\mathcal{G}_{K}^{(p)}$ counts $p^K$ nodes and any pair of nodes which occur to first be connected at the $d$-th iteration displays a coupling
\begin{eqnarray}\label{HPScoupling}
J^{(p)}(d,K,\sigma)&=&\sum_{l=d}^{K}J^{(p)}(l,K,\sigma)=\nonumber\\
\label{eq:Jdef}
&=&\sum_{l=d}^{K}p^{-l[p-2(1-\sigma)]}=\nonumber\\&=&\frac{p^{(1-d)[p-2(1-\sigma)]}-p^{-K[p-2(1-\sigma)]}}{p^{p-2(1-\sigma)}-1}=\\
&=& \frac{[J^{(p)}(1,1,\sigma)]^d - [J^{(p)}(1,1,\sigma)]^{K+1}}{1 - J^{(p)}(1,1,\sigma)} .
\end{eqnarray}
The iterative procedure is summarized in Fig.~\ref{fig:G}.
\newline
Remarkably, this procedure allows for a definition of metric:  two nodes $i$ and $j$ in the set of nodes $\mathcal{V}$ are said to be at distance $d_{ij}=d$ if they occur to be first connected at the $d$-th iteration. Such a distance is ultrametric, in fact, it fulfils the following conditions: $i)$ positivity, namely, $d_{ij}\geq 0$ $\forall i,j\in \mathcal{V}$ and $d_{ij}=0$ if and only if $i=j$; $ii)$ symmetry, namely, $d_{ij}=d_{ji}$ $\forall i,j\in \mathcal{V}$; $iii)$ ultrametric inequality, namely, $d_{ij}\leq\max(d_{ik},d_{kj})$ $\forall i,j,k\in \mathcal{V}$. 
\newline
In general, for a given node, the total number of neighbours at distance $d$ is $(p-1)p^{d-1}$.

The resulting weighted graph $\mathcal{G}^{(p)}_K$ (simply referred to as $\mathcal{G}^{(p)}$ to lighten the notation) is undirected and fully connected. Its nodes make up a set $\mathcal{V}$ of size $N=p^K$ and are labeled as $i=1,...,N$. The pattern of couplings can be encoded by a $N \times N$ matrix $\mathbf{J}^{(p)}$, whose entry $J_{ij}^{(p)}$ depends on the couple $(i,j)$ only through their distance $d_{ij}$, 
being $J_{ij}^{(p)} = J^{(p)}(d_{ij},K,\sigma)$. 

The maximum and the minimum of the weights associated to links are, respectively,
\begin{eqnarray}
\max_{i,j \in \mathcal{V}} J_{ij}^{(p)} = J^{(p)}(1,K,\sigma)&=&\frac{1-p^{-K[p-2(1-\sigma)]}}{p^{p-2(1-\sigma)}-1},\label{J1}\\
\min_{i,j \in \mathcal{V}} J_{ij}^{(p)} = J^{(p)}(K,K,\sigma)&=& p^{-K[p-2(1-\sigma)]}\label{JK} .
\end{eqnarray}

\begin{figure}[tb]
\includegraphics[width=10cm]{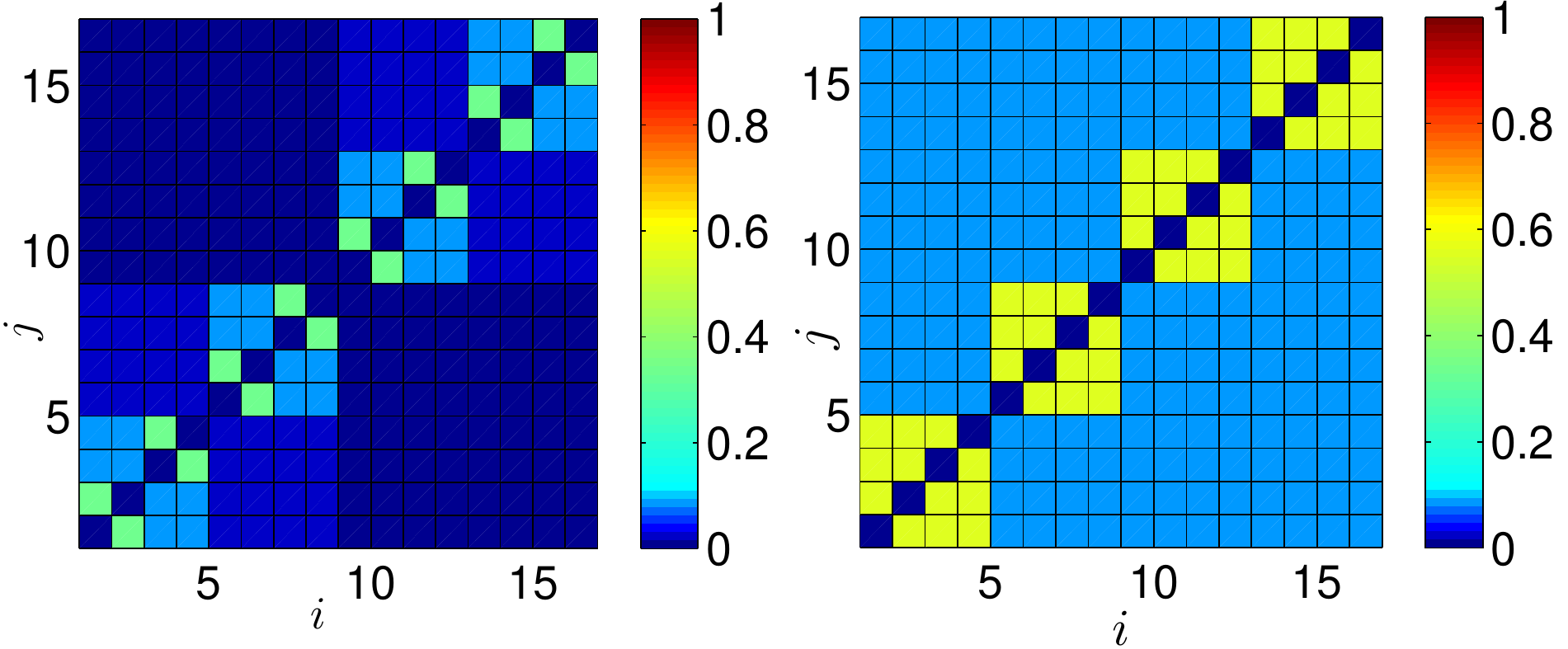}
\caption{(Color
	online) Representation of the coupling matrix $\mathbf{J}^{(p)}$ for the cases $p=2$, $K=4$, $\sigma=1$ (left) and $p=4$, $K=2$, $\sigma=1$ (right). The related networks $\mathcal{G}^{(2)}_4$ and $\mathcal{G}^{(4)}_2$ result of the same size $N=16$, but their structures are not equal, since they are built up following different laws.}\label{fig:J}
\end{figure}

Moreover, by construction, $\mathcal{G}^{(p)}$ turns out to be modular \cite{ABGGTT-PRE2015}, where, with ``modularity'' we mean  
the capability of the graph to be divided into modules (clusters or communities): high modularity means that there are strong (or dense) connections between elements belonging to the same module, and weak (or sparse) interconnections between different modules; conversely, low modularity means that the weights (or the links themselves) are distributed homogeneously. Indeed, in $\mathcal{G}^{(p)}$, at the highest level of resolution, modules are constituted by $p^{K-1}$ cliques of size $p$; at the next step of resolution modules are $p^{K-2}$ cliques of size $p^2$, and so on for further steps. 
Two examples of coupling matrix for different values of $p$ are shown in Fig.~\ref{fig:J}: in both cases the ultrametric pattern as well as the modular structure are evident.


Now, we can look at the nodes of $\mathcal{G}^{(p)}$ as the states of a Markov process, characterized by transition rates from state $i$ to state $j$ given by $J_{ij}^{(p)}$ (under proper normalization). This basically corresponds to a biased random walker on a complete graph endowed with an ultrametric distance where the bias favours closer nodes. Thus, this kind of investigation may shed light on the role of a metric 
%
 %
 as for, e.g., diffusion-reaction processes, traffic models, epidemics \cite{Barrat-2008,FirstPassage-Book2014,Zhang-PRE2013}.
%
%
More precisely, we introduce the (doubly) stochastic matrix $\mathbf{P}^{(p)}$, whose entry $P^{(p)}_{ij}$ represents the probability that the process moves from one state $i$ to a state $j$ at distance $d_{ij}$ and it is defined as 
\begin{equation} \label{eq:p}
P_{ij}^{(p)} \equiv \frac{J_{ij}^{(p)}}{w_i^{(p)}(K,\sigma)}, 
\end{equation}
where $w_i^{(p)}(K,\sigma) \equiv \sum_{j=1, j\neq i}^N J_{ij}^{(p)}$ is the weighted degree of the node $i$, which turns out to be
\begin{eqnarray}
w_i^{(p)}(K,\sigma) &\equiv& 
\frac{N^{-(p-2+2\sigma)}[N(1-pp^{\prime})-1-p^{\prime}]+p^{\prime}(p-1)}{N^{p-2+\sigma}p^{\prime}(p-1)+pN(1-p^{\prime})+p^{\prime}-p}
 \xrightarrow[K \rightarrow \infty]{} \frac{p^{\prime}(p-1)}{(p^{\prime})^2+p^{\prime}(p-1)+p},
\end{eqnarray}
%
%
where $p^{\prime} = p^{p-2(1-\sigma)}$ and $N=p^K$; notice that 
, due to the homogeneity of $\mathcal{G}^{(p)}$,  $w_i^{(p)}(K,\sigma)$ is actually site-independent, hence hereafter we will drop the index $i$. Thus, $P_{ij}^{(p)}$ depends on the couple $(i,j)$ only through its distance $d_{ij}$ and, with a little abuse of notation, we can therefore ``coarsen''  the matrix $\mathbf{P}^{(p)}$ into the (discrete) function $P^{(p)}(d,K,\sigma)$ representing the probability that the system moves from one arbitrary state to another state at distance $d$, that is,
 \begin{equation} \label{eq:prob}
 	P^{(p)}(d,K,\sigma)=\frac{J^{(p)}(d,K,\sigma)}{w^{(p)}(K,\sigma)} =
 	\frac{(N^{p-2+2\sigma}{p^{\prime}}^{(1-d)}-1)(p^{\prime}-p)}{p^{\prime}(p-1)N^{p-2+2\sigma}+pN(1-p^{\prime})+p^{\prime}-p}\xrightarrow[N \rightarrow \infty]{}\frac{(p^{\prime}-p)(p^{\prime})^{1-d}}{p^{\prime}(p-1)}.
 \end{equation} 
 Of course, 
 \begin{eqnarray}
 \max_{i,j \in V} {P}^{(p)}_{ij}(K,\sigma)& =& \max_d {P}^{(p)}(d,K,\sigma) = {P}^{(p)}(1,K,\sigma) =\nonumber\\
 &=&
 \frac{(N^{p-2+2\sigma}-1)(p^{\prime}-p)}{p^{\prime}(p-1)N^{p-2+2\sigma}+pN+2p-p^{\prime}}\label{p1}\\ 
 \min_{i,j \in V} {P}^{(p)}_{ij}(K,\sigma) &=& \min_d {P}^{(p)}(d,K,\sigma) = {P}^{(p)}(K,K,\sigma) = \nonumber\\
&=& 
\frac{(p^{\prime}-p)(p^{\prime}-1)}{p^{\prime}(p-1)N^{p-2+2\sigma}+pN(1-p^{\prime})+p^{\prime}-p}.
\label{pk}
 \end{eqnarray}

Moreover, recalling that the total number of nodes at distance $d$ from a generic site  is equal to $(p-1)p^{d-1}$, the probability $\tilde{P}^{(p)}(d, K,\sigma) $ to jump to \emph{any} site at distance $d$ from a given starting point, say $i$, is 
 \begin{eqnarray}
 	\tilde{P}^{(p)}(d, K,\sigma) &=& \sum_{V \ni j : d_{ij}=d}{P}^{(p)}_{ij}(K,\sigma) 
 	= (p-1)p^{d-1} {P}^{(p)}(d,K,\sigma) =\nonumber\\
 	&=& \frac{(p-1)p^{d-1}J^{(p)}(d,K,\sigma)}{w^{(p)}(K,\sigma)} = \nonumber\\
 	&=&
 	\frac{(p-1)(p-p^{\prime})\Big[(p/p^{\prime})^{d-1}N^{p-2+2\sigma}-p^{d-1}\Big]}{p^{\prime}(p-1)N^{p-2+2\sigma}+pN(1-p^{\prime})+p^{\prime}-p}
  \xrightarrow[N \rightarrow \infty]{} \frac{p^{d-1}(p^{\prime}-p)}{(p^{\prime})^{d}}.
  \end{eqnarray} 

\begin{figure}[h!]
	\includegraphics[width=13cm]{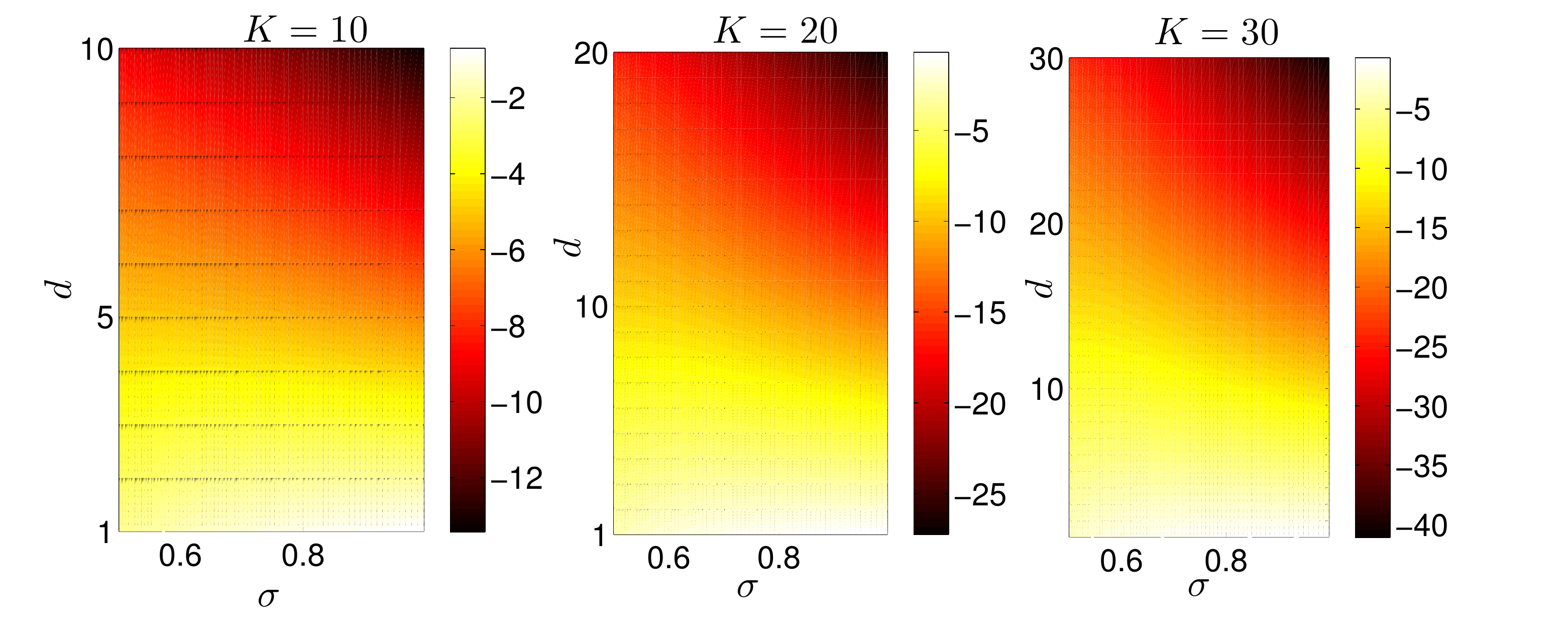}
	\caption{(Color
		online) In order to highlight how $\sigma$ and $K$ affect the degree of homogeneity of the network $\mathcal{G}_K^{(p)}$, we plot $\log_2(P^{(2)}(d,K,\sigma))$, evaluated according to Eq.~\ref{eq:p}, as a function of $d\in[1,K]$ and $\sigma\in(\frac{1}{2},1]$, respectively, fixing $K=10,20,30$. Notice that, for relatively small values of $d$, by increasing $\sigma$ the probability $P^{(2)}(d,K,\sigma)$ also grows, while, for relatively large values of $d$, by increasing $\sigma$ the probability $P^{(2)}(d,K,\sigma)$ also decreases. }\label{fig:trendP}
\end{figure}

We conclude this section with a remark which will be useful when computing mean first passage times (see Sec.~\ref{sec:hitting2}  and Sec.~\ref{sec:hittingp}).
Let us consider $P^{(p)}(1,K,\sigma)$ and $P^{(p)}(K,K,\sigma)$ (see Eq~\ref{p1} and \ref{pk}): for any given $\sigma$, they both decrease with $K$, as expected since the support of the distribution gets larger while normalization must be preserved; conversely, if we fix $K$ and we let $\sigma$ vary in the interval $(\frac{1}{2},1]$, we see that $P^{(p)}(1,K,\sigma)$ is a monotonic increasing function while $P^{(p)}(K,K,\sigma)$ is a monotonic decreasing function. This means that, when $\sigma$ is large, the pattern of transition probabilities is more inhomogeneous and ``close'' nodes are more likely to be reached in a single jump, while, when $\sigma$ is small, the system tends to be more homogeneous, see also Fig.~\ref{fig:trendP}.

\section{Asymptotic properties of $\mathcal{G}^{(2)}$} \label{sec:asymptotics2}


Let us consider  a stochastic process $\{X_n\}$ defined in $\mathcal{G}^{(2)}$ and such that, at any time step $n$, $X_n$ can assume a value in $[1,...,N]$. 
The probability for this system to move from state $X_{n} = i$ to $X_{n+1} = j$ is given by $P_{ij}^{(2)}= P^{(2)}(d_{ij},K,\sigma)$ (see Eq.~\ref{eq:p}).
As highlighted by the expression in Eq.~\ref{eq:prob}, $P^{(2)}(d,K,\sigma)$ is decreasing exponentially with the distance $d$ and, as remarked in the previous section, the decrease is amplified when $\sigma$ is large.  In particular, for large sizes (i.e., $K \gg 1$), $P^{(2)}(K,K,\sigma)/P^{(2)}(1,K,\sigma) \approx 2^{-2 K \sigma}$. 
\newline
However, if one looks at the probability $\tilde{P}^{(2)}(d,K,\sigma)$ to reach \emph{any} site at distance $d$, the fast decrease of $P^{(2)}(d,K,\sigma)$ with respect to $d$ is smoothened by the exponential growth of the number of nodes at distance $d$ from the starting one. 
We therefore expect an interplay between $d$ and $\sigma$ that selects which class of nodes (either the ``close'' or the ``distant'' ones) is more likely to be reached. This can be obtained by solving the following inequality
 \begin{equation}
 	\tilde{P}^{(2)}(d,K,\sigma) = \sum_{l=1}^d 2^{l-1}P^{(2)}(l,K,\sigma)>\sum_{l=d+1}^{K}2^{l-1}P^{(2)}(l,K,\sigma) = 1 - \tilde{P}^{(2)}(d,K,\sigma),\label{animaletti1}
 \end{equation}
 where in the left hand side of the inequality we have the total probability to move within a distance $d$, while in the right hand side we have the probability to overcome the distance $d$. In particular, in the limit $K\rightarrow\infty$, we get
  \begin{eqnarray}
  	\lim_{K\rightarrow\infty} \left[ \sum_{l=1}^d 2^{l-1}P^{(2)}(l,K,\sigma) \right] &>&\lim_{K\rightarrow\infty} \left[ \sum_{l=d+1}^{K}2^{l-1}P^{(2)}(l,K,\sigma)\right]\label{pd}\\
  	\Rightarrow 1-2^{d-2 d \sigma }&>&2^{d-2 d \sigma }  \nonumber\\
  	\Rightarrow \sigma&>&\frac{d+1}{2d}.\label{sigmas}
  \end{eqnarray}
Thus, the Markov process is more likely to jump within a distance $d$, as long as we fix $\sigma$ large enough (according to the value of $d$). On the other hand, for a given $\sigma$, nodes within a distance $\lceil 1/(2 \sigma - 1) \rceil$ are more likely to be reached (and, a fortiori, this holds for any larger distance $d> \lceil 1/(2 \sigma - 1) \rceil$). In particular, when $\sigma=1$, one gets that the inequality (\ref{pd}) holds for any $d>1$, while  if $\sigma\rightarrow 1/2$ the threshold for $d$ tends to increase, consistently with the growing homogeneity of the underlying network. 

These results nicely match with those obtained in \cite{ABGGTT-NN2015}, where, for the Dyson ferromagnetic model, the stability of  spin configurations pertaining to a small module corresponding to $\mathcal{G}^{(2)}_{k>1}$ is shown not to be influenced by the behavior of spins of different modules.

Also, it is worth noticing that, as long as $d$ is finite, none of the probabilities in the inequality (\ref{pd}) is exactly equal to $1$ in the infinite size limit (i.e., for $K\rightarrow\infty$), that is, there is no certainty that a stochastic process surely jumps to a ``near" state, or surely jumps to a ``distant'' state.
On the other hand, if we consider $d=d(K)$, Eq.~\ref{pd} can be rewritten as
\begin{eqnarray}
	\lim_{K\rightarrow +\infty}\sum_{l=1}^{d(K)}2^{l-1}P^{(2)}(l,K,\sigma) &>&	\lim_{K\rightarrow +\infty}\sum_{l=d(K)+1}^{K}2^{l-1}P^{(2)}(l,K,\sigma)\nonumber\\
	\lim_{K\rightarrow +\infty}\frac{4^{\sigma}-2+(2N)^{2\sigma}-2^{d(K)}(4^{\sigma}-2+(2N)^{2\sigma}2^{-2\sigma d(K)})}{(2N)^{2\sigma}+(2N-1)(2^{2\sigma}+1)} &>& \lim_{K\rightarrow +\infty}\frac{2^{d(K)}(4^{\sigma}-2+(2N)^{2\sigma}2^{-2\sigma d(K)})-2N(4^{\sigma}-1)}{(2N)^{2\sigma}+(2N-1)(2^{2\sigma}+1)},\nonumber\label{ineq}\\
\end{eqnarray}
and, by requiring $ d(K) \underset{K \rightarrow \infty} {\longrightarrow} \infty$, we get that the left hand side of the inequality (\ref{ineq}) tends to $1$, while, of course, the right one tends to $0$. Therefore, in the limit of infinite size, the process can not reach the largest distance $K$, but it remains constrained in an infinite sized subgraph $\mathcal{G}^{(2)}_{d(K)}$. This holds in the case $\displaystyle \lim_{K\rightarrow +\infty}d(K)/K=\beta$, with $\beta\in(0,1]$, including, for instance, $d(K)=K-l$ (with $l$ finite), or $d(K)=\alpha K$ (with $0 < \alpha<1$), and also in the case $\displaystyle \lim_{K\rightarrow +\infty} d(K) / K =0$, including, for instance, $d(K)=\log(K)$.

In order to further investigate whether the pattern for couplings $\mathbf{J}^{(2)}$ does induce any ``trap'' effect for the Markov process, which would be then confined in the neighbourhood of the starting node, we consider two extreme cases, namely we evaluate $i.$ the mean time to first escape from the starting dimer (i.e., a couple of nodes at distance $1$, namely the subgraph corresponding to $\mathcal{G}^{(2)}_{1}$) and $ii.$ the mean time to first escape from the starting main branch (i.e., the subgraph corresponding to $\mathcal{G}^{(2)}_{K-1}$).

Let us start with the former and consider a couple of nodes $i$ and $j$ such that $d_{ij}=1$, and let us compute the probability  for a stochastic process to eventually escape from this ``dimer". Of course, due to the symmetry of this subgraph, $P_{ij}^{(2)} = P_{ji}^{(2)}$ and both equal $P^{(2)}(1,K,\sigma)$, here denoted with $q_K$ for simplicity. 
Then, the probability to bounce between $i$ and $j$ for $n$ time steps, without ever leaving this original dimer is 
\be
q_K^n = \Big[\frac{\left(4^\sigma-2\right) \left(N^{2 \sigma}-1\right)}{2^{2\sigma}(N^{2\sigma}+1)+2N(1-2^{2\sigma})-2}\Big]^n.
\ee
This probability is asymptotically vanishing regardless of whether we first let $n \rightarrow \infty$ and then take  $K \rightarrow \infty$ (this case is trivial since $q_K <1, \forall K>0$), or we first let $K \rightarrow \infty$ and then take  $n \rightarrow \infty$:
\begin{eqnarray}
\lim_{K\rightarrow\infty}\lim_{n\rightarrow\infty}q_K^n &= & 0,\\
\lim_{n\rightarrow\infty}\lim_{K\rightarrow\infty}q_K^n &=&
\lim_{n\rightarrow\infty}\Big(\frac{4^{\sigma}-2}{4^{\sigma}}\Big)^n=0.
\end{eqnarray}
Moreover, the mean time $t(1)$ necessary to first escape from the original dimer is 
\begin{eqnarray}
t(1) &=&\sum_{n=1}^{+\infty}n q_K^{n-1}(1-q_K)=\frac{2^{2\sigma}(2N)^{2\sigma}+2N(1-2^{2\sigma})+2^{2\sigma}-2}{2[(2N)^{2\sigma}+N(1-2^{2\sigma})+2^{2\sigma}-2]} 
{\underset{N \to \infty}\longrightarrow}  2^{2 \sigma-1}.
\end{eqnarray}
As expected, $t(1)$ increases with $\sigma$, consistently with the fact that $P(1,K,\sigma)$ is monotonically increasing with $\sigma$; also, $t(1)$ remains finite for any choice of the parameters $\sigma$ and $K$.

Let us now evaluate the probability that the process eventually arrives at distance $K$, i.e., it escapes from the starting subgraph $\mathcal{G}_{K-1}^{(2)}$. To this aim it is convenient to look at $\mathcal{G}_K^{(2)}$ as a bipartite graph, where the sites belonging to the two largest modules are coalesced into two ``super nodes''. The process is therefore described by a sequence of random variables $X_n=\{0,1\}$ such that when $X_n=0$ the process is in the starting subgraph $\mathcal{G}^{(2)}_{K-1}$, while if $X_n=1$, it is in the complementary subgraph. At the initial time step $X_0=0$ and for the successive steps the probability that $X_n = X_{n+1}$ is $1-2^{K-1}P^{(2)}(K,K,\sigma)$, denoted with $f_K$, while the probability that 
$X_n \neq X_{n+1}$ is $2^{K-1}P^{(2)}(K,K,\sigma) = 1 -f_K$.
The probability to stay for $n$ steps in the same branch of the graph is
\begin{equation}
	f_K^n=[1-2^{K-1}P(K,K,\sigma)]^n \approx
	 e^{-n\frac{N^{(1-2\sigma)}}{2}},
\end{equation}
whose limits are
 \begin{eqnarray}
 	\lim_{K\rightarrow+\infty}\lim_{n\rightarrow+\infty} f_K^n =0,\\
 	\lim_{n\rightarrow+\infty}\lim_{K\rightarrow+\infty} f_K^n =1.
 \end{eqnarray}
The non-commutability of the limits suggests that the distance dependent couplings encoded by $\mathbf{J}^{(2)}$ induce the breakdown of ergodicity in the thermodynamic limit. As shown in \cite{ABGGTT-PRL2015,ABGGTT-PRE2015} this has also important effects on the thermodynamic performances of the network. 
 \newline
 The mean time $t(K)$ to first reach the opposite branch of the network is 
 \begin{eqnarray}
 	t(K) =\sum_{n=1}^{\infty} n  f(1-f_K)^{n-1} \nonumber= \frac{1}{f_K} {\underset{K \to \infty}\longrightarrow}  \infty,
 \end{eqnarray}
 and for finite size systems it grows exponentially with the size.

\section{Splitting probabilities on $\mathcal{G}^{(2)}$} \label{sec:splitting}

As shown in the previous section, by tuning $\sigma$ we can modulate the degree of homogeneity of the pattern of weights with qualitative consequences on the probability to reach far ($d \sim K$) and close ($d=1$) nodes. Here, we provide another perspective to quantify how $\sigma$ controls the reachability of distant nodes, namely, we look at the splitting probability $P(1|K)$ to first reach the single node at distance $1$ without ever visiting \emph{any} node at distance $K$; $P(K|1)$ is similarly defined as the probability to first reach \emph{any} node at distance $K$ without ever visiting $1$.
For these quantities, we expect the emergence of a critical value $\sigma_{c}$ (possibly depending on $K$) such that, for $\sigma < \sigma_{c}$, the large number of nodes at distance $K$ prevails and $P(K|1)/P(1|K)>1$, while for $\sigma > \sigma_{c}$, the inhomogeneity of the pattern of weight prevails and $P(K|1)/P(1|K)<1$.
 
\begin{figure}[h!]
 	\includegraphics[width=6cm]{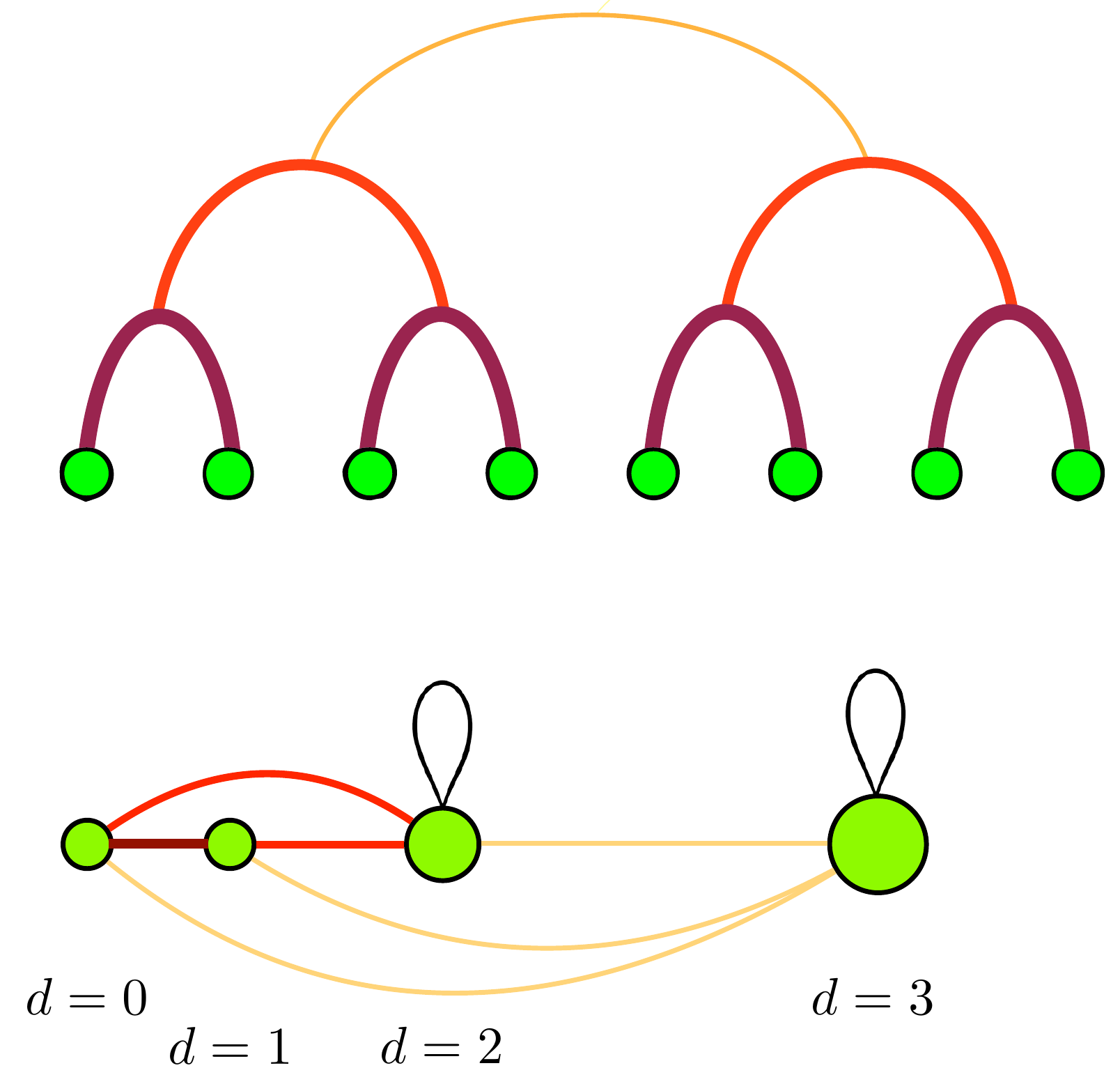}
 	\caption{(Color
 		online) Upper panel: representation of $\mathcal{G}^{(2)}$ for $K=3$, where the thickness and the darkness of links mirror the magnitude of the related coupling. Lower panel: fixing the leftmost node as starting node, all the nodes sharing the same distance have been collapsed into a unique ``super node'' whose size mirrors the number of nodes collapsed; this number scales with the distance as $2^{d-1}$. Notice that super nodes at distance $d>1$ display a probability to remain on the same node, represented by a self loop.  Again, the thickness and the darkness of links mirror the magnitude of the related coupling.}\label{fig:collapse}
 \end{figure}

Again, we consider a Markov chain defined on $\mathcal{G}^{(2)}$ and characterised by the transition probability $P^{(2)}(d,K, \sigma)$ (see Eq.~\ref{eq:p}). Since the nodes at the same distance from $i$ are indistinguishable, we can consider a simplified version of  $\mathcal{G}^{(2)}$ where equivalent nodes are collapsed (see Fig.~\ref{fig:collapse}). The resulting graph has size $K+1$, accounting for nodes which are at distance $1, 2, ..., K$ from $i$ and $i$ itself ($d_{ii}=0$). 

In this way we build up a  $(K+1)\times(K+1)$ transition matrix  $\mathbf{M}$ whose entry $M_{lm}$ is the probability to jump from a node $j$, such that $d_{ij}=l$, to another node $k$, such that $d_{ki}=m$ \footnote{Notice that, in order to keep the notation simple, in $M_{lm}$ the indices actually run from $0$ to $K$, which is the range of possible distances among nodes.}. We stress that $\mathbf{M}$ never coincides with the transition matrix $\mathbf{P}^{(2)}$ of the network, since the elements of $\mathbf{M}$ are states of a random variable whose values are distances from a generic starting node, and not the nodes of the graph itself. This formalization permits to deal with transition matrices with smaller dimensions (i.e., $(K+1)\times (K+1)$, rather than $2^K \times 2^K$), maintaining a full consistency with the model. We can now fix  $\mathcal{T}=\{0,2,...,K-1\}$ the set of transient states, such that $|\mathcal{T}|=K-1$, and $\mathcal{A}=\{1,K\}$ the set of absorbing states, such that $|\mathcal{A}|=2$. If we re-order the columns and the rows in a such a way that the transient states come first, we can write the following  canonical form for $\mathbf{M}$:
\begin{equation}
\mathbf{M}=\left (
\begin{array}{c|c}
\mathbf{Q} & \mathbf{R} \\
\hline
\mathbf{0} & \mathbf{I}
\end{array}
\right ),\label{matrixM}
\end{equation}
where the block $\mathbf{Q}$ is the transition $(K-1)\times (K-1)$ submatrix whose elements are the probabilities to jump from a transient state to another transient state, while $\mathbf{R}$ is $(K-1)\times 2$ submatrix  whose entries are the probabilities to jump from a transient state to an absorbing one. Also, $\mathbf{I}$ is a $2\times 2$ identity matrix representing the two absorbing states, while $\mathbf{0}$ is a $2\times (K-1)$ zero matrix, that accounts for the fact that it is impossible to escape from the absorbing states. One can see that the probability that the process is eventually absorbed is $1$ for all $K$, that is $\lim_{n\to\infty} \mathbf{Q}^n=\mathbf{0}$, since the entries $Q_{ij}<1$, $\forall i,j=1,...K-1$, with $K<+\infty$.  Moreover, for an absorbing Markov chain, the matrix $\mathbf{I-Q}$ has an inverse $\mathbf{N}$, whose entries are the expected number of times the process arrives in state $j$ if it starts in state $i$. Our goal is to determine the matrix $\mathbf{E}$ defined as $\mathbf{E}=\mathbf{NR}$ and of size $(K-1)\times 2$, where the elements $E_{ij}$ are the probabilities that the stochastic process is absorbed in state $j$ if it starts in transient state $i$. 
The splitting probabilities considered here correspond, respectively, to $E_{i1}=P(1|K)$ and to $E_{i2}=P(K|1)$, where $i$ is the starting generic node. 

\begin{figure}[h!]
 	\includegraphics[width=10cm]{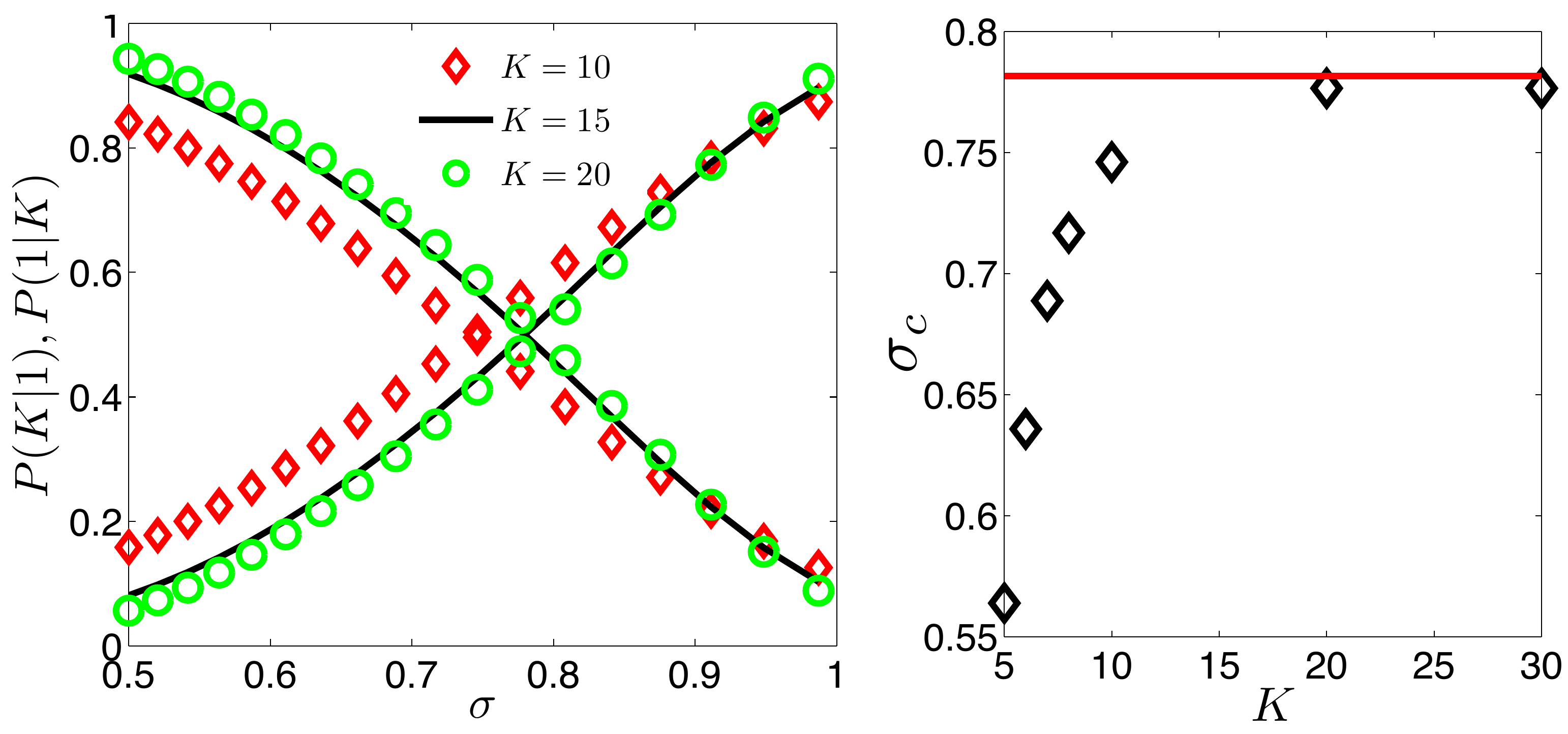}
	 	\caption{(Color
	 		online) Left panel: Splitting probabilities $P(K|1)$ and $P(1|K)$ of being absorbed, respectively, at distance $K$ and at distance $1$, as functions of $\sigma$ and for different values of $K$ (i.e., $K=10,15,20$, as shown in the legend). Data are obtained via numerical simulations, with average over $10^3$ realizations. The point where the curves intersect corresponds to $\sigma_{c}$ and when $\sigma > \sigma_{c}$ it is easier to jump to distant nodes rather than closer nodes. Right panel: the values of $\sigma_{c}$ extracted in this way are plotted as a function of $K$. Notice that $\sigma_c$ grows with $K$, possibly reaching an asymptotic value $\approx 0.8$ represented by the solid line. }\label{fig:criAsympt}
 \end{figure}

In Appendix \ref{sec:b} we show some algebraic passages which allows semplifying the expression for $\mathbf{E}$ in such a way that its numerical calculation is fastened. Having an estimate for $P(K|1)$ and for $P(1|K)$, we focus on the ratio
\begin{equation} \label{eq:ratio}
r(K,\sigma)=\frac{P(K|1)}{P(1|K)}, 
\end{equation} 
and we study its behavior as a function of $\sigma$ when $K$ is fixed. As can be inferred from Fig.~\ref{fig:criAsympt}, the probability $P(1|K)$ of eventually reaching the node at distance $1$ without ever reaching any node at distance $K$ is larger than its complementary $P(K|1)$ (i.e., $r(K, \sigma)<1$) provided that $\sigma$ is large enough, how large depending on the size $K$. The threshold value, referred to as $\sigma_{c}$, can be estimated using the matrix $\mathbf{E}$ (see also Eq.~\ref{B}) and requiring $E_{i1}=E_{i2}$.
We find that the critical value $\sigma_{c}$ grows with $K$: a large size requires a relatively large degree of inhomogeneity (i.e. large $\sigma$) to compensate the fact that the number of nodes at maximum distance is exponentially growing with $K$.


\section{Hitting times on $\mathcal{G}^{(2)}$} \label{sec:hitting2}
In the previous sections we outlined that in infinite size systems ergodicity can be broken and distant nodes cannot be reached (see Sec.~\ref{sec:asymptotics2}). On the other hand, even in finite-size systems inhomogenity induced by distance can be strong enough (for large values of $\sigma$) that far nodes are unfavoured even if their number is much larger that close nodes (see Sec.~\ref{sec:splitting}). Hence, one would naively expect that if we look at a single node at distance $K$ and compare the mean first passage time $\tau(K)$ to that single node and the mean first passage $\tau(1)$ to the single node at distance $1$, then $\tau(K)/\tau(1)$ must increase fast with the system size. 

Actually, as we are going to show, even in a regime of high inhomogeneity for $\mathcal{G}^{(2)}$ (i.e., for $\sigma$ relatively large), this ratio exhibits a slow growth (if any).
In general, the mean time to first reach a vertex $j$ starting from the vertex $i$ depends on the couple $(i,j)$ only through its distance $d_{ij}$. Thus, posing $d_{ij}=d$, the mean time from $i$ to $j$, referred to as $\tau(d)$ can be written as 
\begin{eqnarray}\label{eq:system}
\tau(d) &=& P(d)+\sum_{l=1}^{d-1}2^{l-1}P(l)\Big[\tau(d)+1\Big]+\sum_{l=1}^{d-1}2^{l-1}P(d)\Big[\tau(l)+1\Big]+\sum_{l=d+1}^{k}2^{l-1}P(l)\Big[\tau(l)+1\Big].\label{rec}
\end{eqnarray}
We remark that, to lighten the notation, the probability $P(d,K,\sigma)$ to move toward a node at distance $d$ from our starting point is simply denoted as $P(d)$, i.e., $P(d) = P^{(2)}(d,K,\sigma)= J^{(2)}(d,K,\sigma) / w^{(2)}(K,\sigma)$.

\begin{figure}[h!]
	\includegraphics[width=12cm]{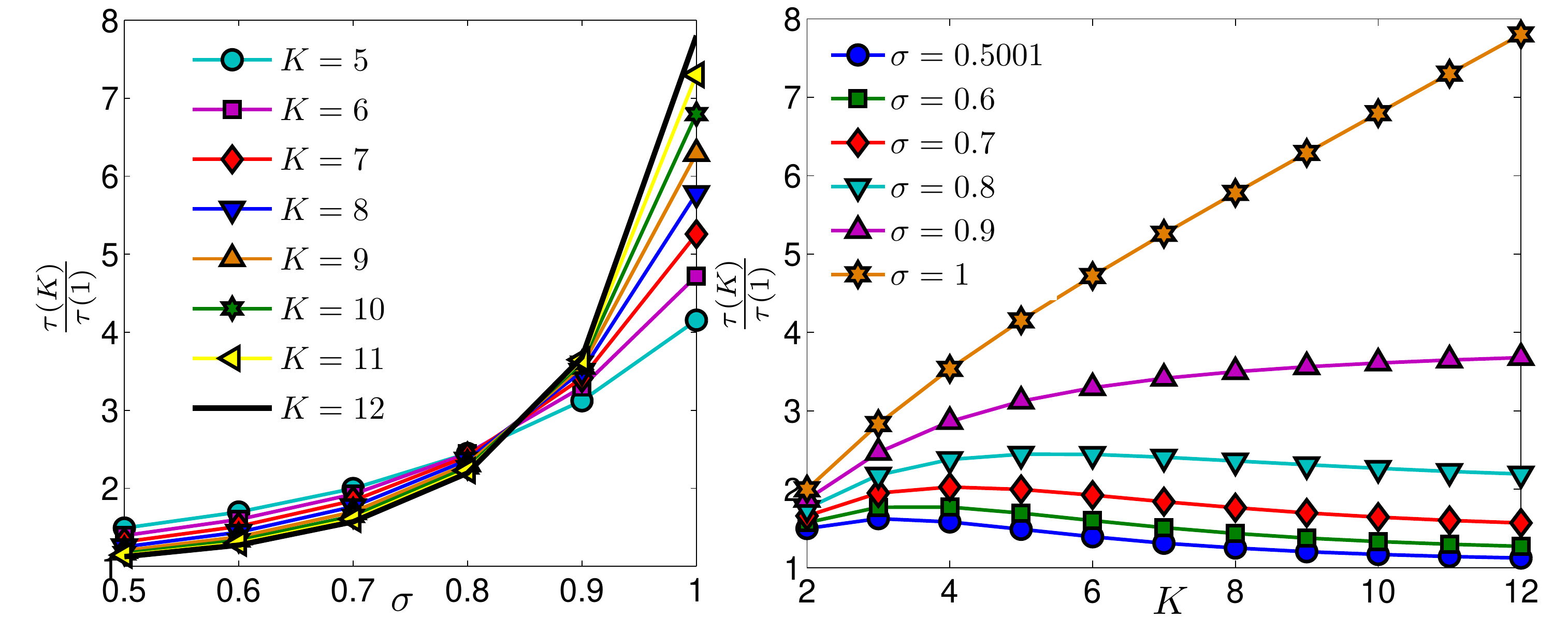}
	\caption{(Color online) Left panel: Trend of the ratio $\tau(K) / \tau(1)$ versus $\sigma\in(0.5,1]$ and for different choices of $K$ as shown in the legend; the rate of growth increases with the size. Right panel: trend of $\tau(K) / \tau(1)$ versus  $K \in [2,12]$ and for different choices of $\sigma$ as shown in the legend. Notice that $\tau(K) / \tau(1)$ exhibits a peak at a value of $K$ which depends monotonically with $\sigma$. The values of $\tau(1)$ and $\tau(K)$ are obtained via the inverse pseudo-Laplacian method.}\label{fig:tKt1}
\end{figure}

Now, Eq.~\ref{eq:system} contains  four contributions to be computed.
The first one represents the probability to move directly from the source $i$ to the target $j$; in the second one the process jumps at the beginning toward one of the nodes $h$ such that $d_{ih}<d_{ij}$ (implying $d_{hj}=d_{ij}$) and then it arrives on the target; the third one describes the probability to jump toward one of the sites at distance $d$ but different from $j$, and then to arrive on $j$; and finally one considers the stochastic process going away from the target toward a node $h$ such that $d_{hj}>d_{ij}$ and then arriving on the target. \\
With some algebraic manipulations we build up the differences between terms for $d+1$ and $d$ and between terms for $d+2$ and $d+1$, obtaining the following expression
\begin{eqnarray}\label{eq:rec}
\tau(d+2)-[1+A(d,K,x)]\tau(d+1) + A(d,K,x)\tau(d)=0,
\end{eqnarray}
where $A(d,K,x) = [x^{K+1-d}-1]/[2(x^{K-1-d}-1)]$, and $x=2^{2\sigma-1}$.
As shown in Appendix \ref{sec:a}, starting from Eq.~\ref{eq:rec}, one can prove that 
\begin{equation}\label{leadingHFM}
\frac{\tau(K)}{\tau(1)}=1+(x-1)\frac{x^K-1}{2^K}\sum_{m=1}^{K-1}\frac{(2x)^m}{(1-x^m)(1-x^{m+1})}.
\end{equation}
By further handling this expression (again details can be found in Appendix \ref{sec:a}), one can see that the behavior of $\frac{\tau(K)}{\tau(1)}$ with respect to $K$ depends qualitatively on $\sigma$. In fact, in the limit of large $K$, when $\sigma =1$ the ratio between times grows linearly with $K$ (i.e., logarithmically with the system size), while for $\sigma<1$ it always reaches an asymptotic value corresponding to $1/(2-x)$. This means that, in order for $\tau(K)$ to scale qualitatively faster than $\tau(1)$ as the size is increased, a very inhomogeneous patter of couplings is required; as long as $\sigma<1$,  $\tau(K)$ and $\tau(1)$ remain comparable.
Basically, when $\sigma$ is small, the pattern of weights is rather homogeneous and the mean time to reach any node\footnote{Indeed, if this holds for $\tau(K)$ then it holds a fortiori for any $\tau(d)$ with $d<K$.} is comparable to $\tau(1)$ similarly to what expected for the (bare) complete graph case (recoverable in the limit $\sigma \rightarrow 1/2$).
On the other hand, even in a regime of rather high heterogeneity (say $\sigma = 0.9$) and for a size $N=2^K \approx O(10^3)$, the time to first reach a particular node at distance $K$ is only $4$ times larger (roughly) than the mean time to reach the closest node, although the related transition probabilities are $P(K,K, \sigma)/ P(1,K, \sigma) \approx N^{-2\sigma}$. 
This picture is corroborated numerically, as shown in Fig.~\ref{fig:tKt1}.



\section{Asymptotic properties on $\mathcal{G}^{(p)}$} \label{sec:asymptoticsp}

In this section we extend the analysis of the asymptotic properties already accomplished on $\mathcal{G}^{(2)}$ (see Sec.~\ref{sec:asymptotics2}) to the general case of  $\mathcal{G}^{(p)}$ (again $\mathcal{G}_K^{(p)}=\mathcal{G}^{(p)}$ holds).
The deterministic modularity characterizing $\mathcal{G}^{(p)}$ allows carrying on the same analysis already explained for the case with $p=2$, but, as we will see, the differences in the nature of couplings give rise to qualitative differences. 

Let us consider a Markov process on $\mathcal{G}^{(p)}$ with transition matrix $\mathbf{P}^{(p)}$ of size $N \times N$ with $N=p^K$, and  such that $P_{ij}^{(p)}=P^{(p)}(d_{ij},K,\sigma)=J^{(p)}(d_{ij},K,\sigma)/w^{(p)}(K,\sigma)$, as defined in $(\ref{HPScoupling})$; as usual, $d_{ij}$ represents the distance between the nodes $i$ and $j$.

Now, let us evaluate under which conditions, if any, the transition probability do favor close nodes, namely the following inequality holds
\begin{equation}
	\tilde{P}^{(p)}(d,K,\sigma) = \sum_{l=1}^{d}(p-1)p^{l-1}P^{(p)}(l,K,\sigma)>\sum_{l=d+1}^{K}(p-1)p^{l-1}P^{(p)}(l,K,\sigma) = 1 - \tilde{P}^{(p)}(d,K,\sigma),\label{ineqP}
\end{equation} 
where $p^{l-1}(p-1)$ is the total number of nodes at distance $l$. In the limit $K\rightarrow\infty$ we get
\begin{eqnarray}
	1-p^{-d(p-3+2\sigma)}&>&p^{-d(p-3+2\sigma)}\nonumber\\
	\Rightarrow \sigma&>&\frac{3d-pd + \log_p(2)}{2d}.\label{critical}
\end{eqnarray}
The previous result provides the interval where a stochastic process living on the graph $\mathcal{G}^{(p)}$ is favoured to reach a site at distance less or equal than $d$. Of course, if we fix $p=2$ we recover the interval already obtained in Sec.~\ref{sec:asymptotics2}. Moreover, for $p\geq3$, the right hand side of the inequality is always (i.e., $\forall d \geq 1$) $<1/2$  in such a way that, for any $\sigma \in 91/2,1]$, jumping to the $p-1$ nearest nodes is still the most likely outcome.

Now, let us estimate the probabilities to stay for $n$ steps at minimum distance. In this general case, where $p\geq 3$, the stochastic process can remain within a distance $1$ from the origin choosing uniformly among a set of $(p-1)$ other sites.
We call $q_K=P^{(p)}(1,K,\sigma)$ the probability to jump to one of the closest nodes, so we can compute the probability to remain up to the time step $n$ in the subset made of the $p$ sites at distance $1$. In this case we have
\begin{equation}
	q_K^n=\left[ \frac{(p^{\prime}-p)(1-N^{p-2+2\sigma})}{(N-1)(p^{\prime}-p)+p^{\prime}N^{p-2+2\sigma}(1-p)}\right]^n.
\end{equation}
 Considering the limit for $K\rightarrow\infty$ and $n\rightarrow +\infty$, we have that

\begin{eqnarray}
\lim_{n\rightarrow\infty }\lim_{K\rightarrow\infty }q_K^n &=&\lim_{n\rightarrow\infty } \Bigg[\frac{\left(p^{\prime}-p\right)}{p^{\prime}(p-1)}\Bigg]^n=0,\\
\lim_{K\rightarrow\infty }\lim_{n\rightarrow\infty }q_K^n &=& 0,
\end{eqnarray}
since $q_K<1$. Therefore, it is not possible to force the process to stay permanently in the same $p$-plet of sites. 
Of course, fixing $p=2$, we recover the same computations of Sec.~\ref{sec:asymptotics2}.
 The time to first escape from a fixed $p$-plet is
 
 \begin{equation}
 	t(1)=\sum_{n=1}^{+\infty}nq_K^{n-1}(1-q_K)=\frac{(p^{\prime}p-1)N^{p-2+2\sigma}+pN(p^{\prime}+1)+p(p^{\prime}-1)}{(pp^{\prime}-2p^{\prime}-p)N^{p-2+2\sigma}+pN+2(p^{\prime}-p)}\xrightarrow[N \rightarrow \infty]{}\frac{pp^{\prime}-1}{pp^{\prime}-2p^{\prime}-p}.
 \end{equation}

Analogously to the case $p=2$, $t(1)$ increases as $\sigma$ increases, consistently with the fact that $P^{(p)}(K,1,\sigma)$ is monotonically increasing with respect to $\sigma$; also, $t(1)$ remains finite regardless of the parameters $\sigma$ and $K$.

We can also look at $\mathcal{G}^{(p)}$ as a network composed by $p$ modules at distance $K$ from each other and evaluate the probability to stay for $n$ steps in one of the $p$ communities. This can be formalized as already shown for the case with $p=2$ (see Sec.~\ref{sec:asymptotics2}): we consider a sequence of random variables $X_n=\{0,1\}$ such that when $X_n=0$ the process is in the original community, while if $X_n=1$, it is in one of the sites of the other $p-1$ branches. Denoting with $f_K$ the probability that $X_{n+1}=X_n$, namely $f_K = 1-(p-1)p^{K-1}P^{(p)}(K,K,\sigma)$ and, accordingly, with $1-f_K$ the probability that $X_{n+1}\neq X_n$, we can write the probability to stay for $n$ steps in the same community as
\begin{equation}
	f_K^n=[1-(p-1)p^{K-1}P^{(p)}(K,K,\sigma)]^n\approx e^{-n\frac{p-1}{p}NP^{(p)}(K,K,\sigma)}.
\end{equation}
Computing the limits for $K\rightarrow +\infty$ and $n\rightarrow +\infty$ we get

\begin{eqnarray}
	\lim_{K\rightarrow +\infty}\lim_{n\rightarrow +\infty} f_K^n=0,\nonumber\\
	\lim_{n\rightarrow +\infty}\lim_{K\rightarrow +\infty} f_K^n =1,\nonumber
\end{eqnarray}
where we used the fact that $P^{(p)}(K,K,\sigma)$ tends to zero as $K$ tends to infinity for all $p\geq 2$. 
Hence, again, in the limit of infinite size, ergodicity is broken.

\section{Hitting times on $\mathcal{G}^{(p)}$}\label{sec:hittingp}

\begin{figure}[h!]
	\includegraphics[width=12cm]{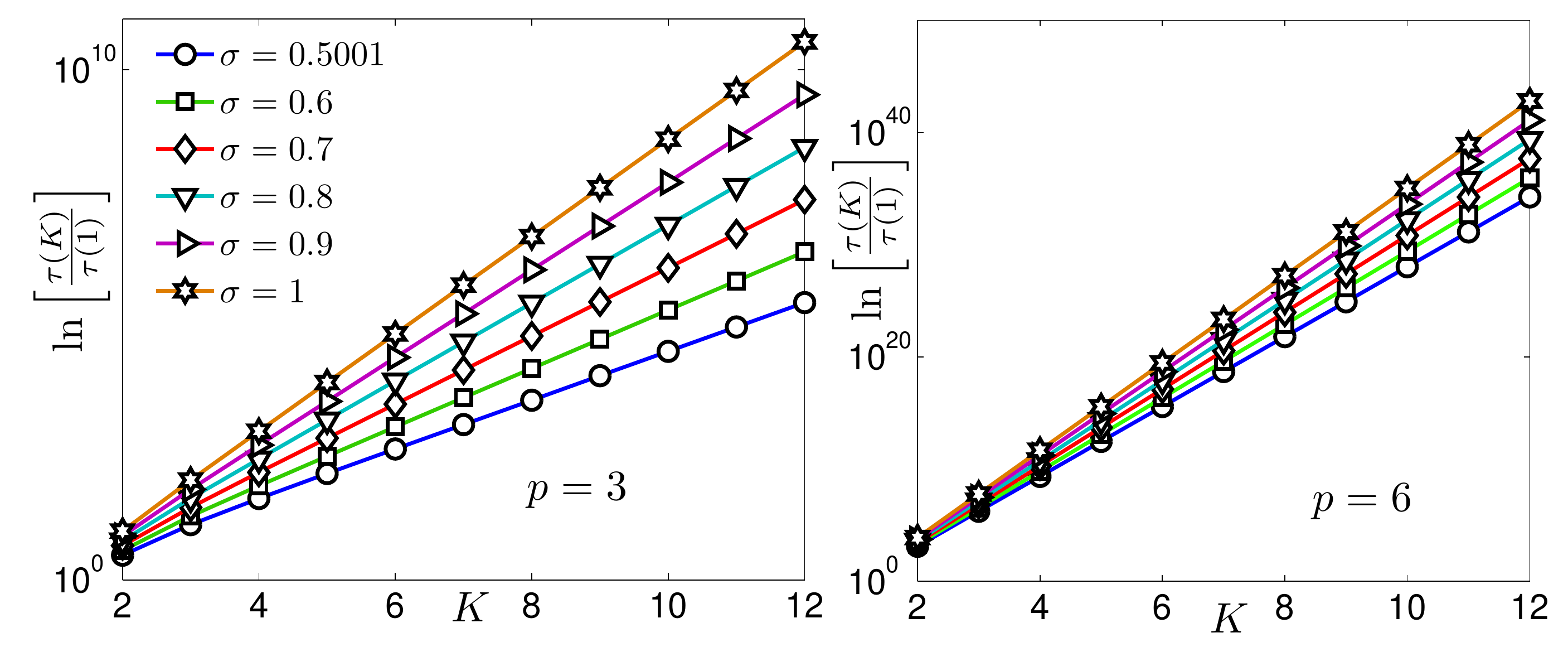}		
	\caption{(Color
		online) Left panel: trend of the ratio $\tau(K)/ \tau(1)$ with respect to $K$ for $p=3$ and for three different values of $\sigma$, as explained by the legend. As one can easily see, there is a strong increasing trend of  this ratio, that grows linearly with respect to $K$ as showed in  Eq. $\ref{tkt1p}$, and that outlines how a stochastic process can not easily reach sites at maximum distance, since the time necessary to arrive there for the first time is extremely large with respect to the time necessary to reach the $p$ nearest neighbors. Right panel: the same analysis was carried on for $p=6$, outlining again the same behavior explained for $p=3$.}\label{fig:sumvsP}
\end{figure}

We now investigate the mean time $\tau(1)$ to first reach a node at distance $1$ and the mean time $\tau(K)$ to first reach a node at distance $K$ on $\mathcal{G}^{(p)}$ with $p\geq 3$.
To this aim we start extending Eq.~\ref{eq:system} as
\begin{eqnarray}
\tau(d)&=& P(d)+(p-2)p^{d-1}P(d) \left [ \tau(d)+1 \right] +
\sum_{l=1}^{d-1}(p-1)p^{l-1}P(l) \left[ \tau(d)+1 \right ] +\sum_{l=1}^{d-1}(p-1)p^{l-1}P(d) \left[ \tau(l)+1 \right] +\nonumber\\
&+&\sum_{l=d+1}^{K}(p-1)p^{l-1}P(l) \left[  \tau(l)+1 \right],
\end{eqnarray}
where, again, we posed $P(d)=P^{(p)}(d,K,\sigma)$.
The contributes of the previous expression are the same as the ones for $\mathcal{G}^{(2)}$, but, in this case, we have to consider that, for example, at the $k$-th level, the network is formed by $p$ communities each formed by $p^{k-1}$ elements. Thus, the stochastic process could make a jump toward a node at distance $d$ that is in a different community than the one where the target is. This fact is accounted for by the term $(p-2)p^{d-1}P(d)[\tau(d)+1]$. 
With some algebraic manipulations we write again a difference equation involving $\tau(d+2)$, $\tau(d+1)$ and $\tau(d)$, obtaining
\begin{eqnarray}\label{recursive3}
\tau(d+2)-[1+B(d,K,x)]\tau(d+1)+B(d,K,x)\tau(d)=0,
\end{eqnarray}
where $B(x,K,d)=(x^{K+1} - x^d)/[p(x^K-x^{d+1})]$, and $x=p^{p-3+2\sigma}$. Now, Eq.~\ref{recursive3} has the same form of ($\ref{eq:rec}$), and we can proceed analogously (see also Appendix \ref{sec:a}), obtaining the following expression for $\tau(K)$
\begin{eqnarray}\label{42}
\tau(K)&=&\tau(1)\left[ 1+(x^{K+1}-x^K)(x^K-1) \right ] \sum_{j=1}^{K-1}\frac{x^jp^{-j}}{(x^j-x^K)(x^j-x^{K+1})}\\
\label{42b}
&=&\tau(1)\left[ 1+(x-1)(x^K-1) \right]\frac{1}{p^K}\sum_{m=1}^{K-1}\frac{(px)^m}{(1-x^m)(1-x^{m+1})},
\end{eqnarray}
where in the last line we reshuffled the sum with proper new indexes.
Proceeding in the same way as we did for the case with $p=2$, we can study the asymptotic behavior of the ratio $\tau(K)/\tau(1)$, arriving to

\be
\frac{\tau(K)}{\tau(1)}\approx 1+C(x,p)N^{p-4+2\sigma},\qquad \sigma\in(1/2,1],\qquad p>2,\qquad\text{ as } K\rightarrow\infty,\label{tkt1p}
\ee
where $C(x,p)$ is a constant depending on the choice of $x$ and $p$.
It is worth noticing that the exponent of $N$ is always positive, due to the restrictions made on $p$ and $\sigma$. This allows us to state that, as $p\geq 3$, the ratio $\tau(K)/\tau(1)$ scales with the size of the system for every choice of $\sigma$ as checked in Fig.~\ref{fig:sumvsP}. Thus, differently from the case $p=2$, no asymptotic value is retained.
This means that, when the size of the modules that compose the network is larger than $2$,  the time necessary to reach nodes at maximum distance from a starting point grows with the system size faster than the time necessary to reach the nearest sites, regardless of the value of $\sigma$.
This is consistent with fact that, in the thermodynamic limit, a stochastic process is, for any choice of $\sigma$, more likely to jump to the closest node (see the first part of Sec.~\ref{sec:asymptoticsp}) despite the number of farthest nodes is much larger.

Before concluding we stress that a qualitative difference in the first-passage features of $\mathcal{G}^{(2)}$ and $\mathcal{G}^{(p)}$ can be highlighted also by looking at the splitting probabilities. In fact, as deepened in Appendix \ref{sec:c}, when $p>2$ the stochastic process is always (i.e., regardless of $\sigma$ and of $K$) more likely to first visit the closest node rather than any node set at the largest distance despite the latter subset grows exponentially with $K$. 

\section{Conclusions} \label{sec:conclusions}
In this work we focused on the influence of the topology of hierarchical, recursively grown, weighted graphs $\mathcal{G}^{(p)}$ on the stochastic processes embedded on them. These  structures provide interesting generalizations for mean-fields models in statistical-mechanics and, in particular, the case $p=2$ corresponds to the Dyson ferromagnetic model, while the case $p>2$ corresponds to systems with $p$-wise interactions (e.g., $p$-spin models). The main feature of this class of networks is the metric induced by their construction:  a couple of nodes are said to be at distance $d$ if they are connected at the $d$-th iteration. In particular, if one performs overall $K$ iterations, links between nodes connected at the first iteration (so at distance $1$) display a higher weight with respect to links between nodes connected at the $K$-th iteration (i.e., at the maximum distance $K$). Further, weights on links are modulated by a tuneable, positive parameter $\sigma$. 

For the stochastic process considered, the probability to move from one state to another state is given (upon proper normalization) by the weight associated to the link connecting the two corresponding nodes. We therefore expect that its behavior crucially depends on $\sigma$ and on the size $p$ of the smallest module.\\
First, we focused on the case $p=2$ and we analyzed the asymptotic properties showing that ergodicity breaks down in the infinite size limit and, in a single jump, as long as $\sigma$ is large enough, ``distant'' nodes are unlikely to be reached despite their number is exponentially larger than that of ``close'' nodes.\\
Further, we dealt with the splitting probabilities to first reach sites at distance $1$ without ever reach sites at distance $K$ (i.e., $P(1|K)$, and vice-versa (i.e., $P(K|1)$). We recovered the existence of a critical value $\sigma_{\text{c}}$ such that $P(K|1)>P(1|K)$ for every $\sigma<\sigma_{\text{c}}$, while $P(1|K)>P(K|1)$  if $\sigma>\sigma_{\text{c}}$. This fact outlines how the pattern of couplings is shaped by $\sigma$: for large values of $\sigma$ the pattern is rather inhomogeneous in such a way that the stochastic process is more likely to reach the nearest neighbors with respect to any node at maximum distance, despite their number is much larger. Conversely, for small values of $\sigma$ the pattern of weights is rather homogeneous and the interaction strength between two sites at minimum distance cannot  compensate the larger number of sites at maximum distance. Interestingly, this feature matches with the findings obtained from a statistical mechanics perspective for the Dyson ferromagnetic model \cite{Castellana_tesi}, where a threshold value for $\sigma$ is found, such that if $\sigma \leq 3/4$ the mean-field approximation is correct, while if $3/4 < \sigma \leq 1$ the model has a non-mean-field behavior.\\
Moreover, we studied the ratio between the hitting time $\tau(K)$ to first reach a single node at the maximum distance $K$ and the hitting time $\tau(1)$ to first reach the nearest neighbor. Also in this case, the role of $\sigma$ is crucial, in fact, despite the fact that $\tau(K)>\tau(1)$ always holds , the lower $\sigma$ and the closer  $\tau(K)/\tau(1)$ to $1$. In this case it emerges the existence of an asymptotic value for $\tau(K)/\tau(1)$ for $\sigma\in(1/2,1)$ while for $\sigma=1$ the ratio scales linearly with $K$.

Finally, we extended these analysis to the general case $\mathcal{G}^{(p)}$, when $p>2$. Again, we studied the asymptotic properties of the stochastic processes living on $\mathcal{G}^{(p)}$, and in this case it turns out that,  in the infinite size limit, nodes within any fixed distance $d \geq 1$ are always (i.e., for any choice of $\sigma$) most likely to be reached, that is to say, inhomogeneity always prevails. Moreover, as long as the size of the graph is finite, the study performed on the splitting probability and hitting times does not highlight the existence of any threshold value of $\sigma$ able to qualitatively affect the behavior of $P(1|K)/P(K|1)$ and $\tau(K)/\tau(1)$, and the logarithm of the latter scales linearly with $K$. 

We can conclude that in $\mathcal{G}^{(p)}$ the interplay between $\sigma$, that rules the interaction strength between nodes, and $p$, that represents the size of the smallest module, can qualitatively change the features of the network. Moreover, by tuning $p$ from $p=2$ to $p>3$, the emerging properties for the stochastic process defined on $\mathcal{G}^{(p)}$ change abruptly, somehow analogously to what happens in the statistical mechanics ferromagnetic (mean-field) models, where the phase transition looses criticality for $p > 2$ (see e.g., \cite{ABC-ROMP2011}).

\section*{Acknowledgments}
The authors are grateful to Adriano Barra for enlightening discussions and interactions.

\appendix

\section{Splitting probabilities on $\mathcal{G}^{(2)}$} \label{sec:b}
Let us consider the transition matrix $\mathbf{M}$ defined in Eq.~\ref{matrixM}: as the size $K$ grows, the number of transient states grows, and, in particular, we can define a recurrence law for the definition of $\mathbf{Q}$. In fact, one has a sequence of matrices $\{\mathbf{Q}\}_K$ of size $(K-1)\times (K-1)$, where  $K=2,...,+\infty$ is the value of the maximum distance of the system, such that 

\begin{eqnarray}
&&\mathbf{Q}_2= 0,\nonumber\qquad \text{for}\mbox{ }K=2,\\
&&\mathbf{Q}_K =\left (
\begin{array}{c|c}
\mathbf{Q}_{K-1} &  \mathbf{b}\\
\hline
\mathbf{c} & \gamma
\end{array}
\right ), \qquad \text{ for }\mbox{ }K>2,
\end{eqnarray}
where $\mathbf{b}$ is a $(K-2)\times 1$ column vector  whose entries are $b_i=2^{(K-1)}P^{(2)}(K,K,\sigma)$, for every $i=1,...,K-2$, $\mathbf{c}$ is a $1\times (K-2)$ row vector whose entries are $c_i=2^{i-1}P^{(2)}(K,K,\sigma)$, with $i=1,...,K-2$, $\gamma=\sum_{l=1}^{K-2}2^{l-1}P^{(2)}(l,K,\sigma)$, and on the top left we have a block matrix of dimension $(K-2)\times (K-2)$ whose entries are the same of the matrix $\mathbf{Q}$ when the generation of the system is $K-1$, but computed with respect of the total number of level $K$. This means that, for example, for $K=3$, $K=4$ and $K=5$ respectively, the matrix $\mathbf{Q}$, has the following form 

\begin{eqnarray}
\mathbf{Q}_3&=&\left (
\begin{array}{cc}
0 & 2P(2,3,\sigma)\\
P(2,3,\sigma) & P(1,3,\sigma)
\end{array}
\right ),\qquad K=3 \nonumber\\ 
\mathbf{Q}_4&=&\left (
\begin{array}{cc|c}
0 & 2P(2,4,\sigma) & 4P(3,4,\sigma)\\
P(2,4,\sigma) & P(1,4,\sigma) & 4P(3,4,\sigma)\\
\hline
P(3,4,\sigma) & 2P(3,4,\sigma) & P(1,4,\sigma)+2P(2,4,\sigma)
\end{array}
\right ),\qquad K=4\nonumber\\
\mathbf{Q}_5&=&\left (
\begin{array}{ccc|c}
0 & 2P(2,5,\sigma) & 4P(3,5,\sigma) & 8P(4,5,\sigma)\\
P(2,5,\sigma) & P(1,5,\sigma) & 4P(3,5,\sigma) & 8P(4,5,\sigma)\\
P(3,5,\sigma) & 2P(3,5,\sigma) & P(1,5,\sigma)+2P(2,5,\sigma) & 8P(4,5,\sigma)\\
\hline
P(4,5,\sigma) & 2P(4,5,\sigma) & 4P(4,5,\sigma) & \sum_{l=1}^3 2^{l-1}P(l,5,\sigma)
\end{array}
\right ), \qquad K=5,\nonumber\\
\end{eqnarray}
where we posed $P(d,K,\sigma)=P^{(2)}(d,K,\sigma)$, (see Eq.~\ref{eq:prob}), dropping the index $p=2$ to simplify the notation.

One can prove that at every iteration $k$, the matrix $\mathbf{Q}_k$ contains a top-left block, equal to the matrix $\mathbf{Q}_{k-1}$, with corresponding entries depending on its effective size. In general, once the structure of  $\{\mathbf{Q}\}_{l}$, $l=1,...,K$ has been defined, we are going to lighten the notation, writing only $\mathbf{Q}$, to refer to the iteration of a fixed $K$. The most important argument is that, for all $K\in\mathbb{N}$, $K\geq 1$, $\mathbf{Q}$ has always the same block form. Using this fact, the computation of the matrix $\mathbf{N}$ becomes easier, since one knows that
\begin{eqnarray}
\mathbf{N}&=&(\mathbf{I}-\mathbf{Q})^{-1}=\left (
\begin{array}{c|c}
\mathbf{I}-\mathbf{Q} &  \mathbf{b}\\
\hline
\mathbf{c} & 1-\gamma
\end{array}
\right )^{-1}=\nonumber\\
&=& \left (
\begin{array}{c|c}
\mathbf{A} & \mathbf{b}\\
\hline
\mathbf{c} & \delta
\end{array}
\right )^{-1}=\nonumber\\
&=& \left ( \begin{array}{c|c}
\mathbf{A}^{-1}+ \delta^{-1} \mathbf{A}^{-1}\mathbf{b}\mathbf{c}\mathbf{A}^{-1} & - \delta^{-1} \mathbf{A}^{-1}\mathbf{b} \\
\hline
- \delta^{-1} \mathbf{c}\mathbf{A^{-1}} & \delta^{-1}
\end{array}
\right )=\nonumber\\
&=&\left( \begin{array}{c|c}
\mathbf{S} & \mathbf{v}\\
\hline
\mathbf{u} & \delta^{-1}
\end{array}
\right )\nonumber \\\label{N}
\end{eqnarray}
where, in the second line, we posed $\mathbf{A} = \mathbf{I}-\mathbf{Q}$, and $\delta = 1 - \gamma$, while, in the last line, we posed $\mathbf{S} =  \mathbf{A}^{-1}+ \delta^{-1} \mathbf{A}^{-1}\mathbf{b}\mathbf{c}\mathbf{A}^{-1}$, $\mathbf{v}= - \delta^{-1} \mathbf{c}\mathbf{A^{-1}} \mathbf{b}$, and $\mathbf{u} = - \delta^{-1} \mathbf{c}\mathbf{A^{-1}}$. In particular, in the third line, we applied the formula for the inversion of block matrices, in fact, $\mathbf{N}$ is still formed by four blocks of size $(K-2)\times (K-2)$,$(K-2)\times 1$, $1\times (K-2)$ and $1\times 1$. Multiplying  $\mathbf{N}$ and $\mathbf{R}$, that has the form
\begin{equation}
\mathbf{R}= \left ( \begin{array}{c|c}
P(1,K,\sigma) & 2^{K-1}P(K,K,\sigma)\\
P(2,K,\sigma) & 2^{K-1}P(K,K,\sigma)\\
\vdots & \vdots\\
P(K-2,K,\sigma) & 2^{K-1}P(K,K,\sigma)\\
\hline\\
P(K-1,K,\sigma) & 2^{K-1}P(K,K,\sigma)
\end{array}
\right ),
\end{equation}
we obtain the following 
\begin{eqnarray}
\mathbf{E}&=&\mathbf{N}\mathbf{R}=\nonumber\\
\nonumber
&=&\left ( \begin{array}{c|c}
\sum_{j=1}^{K-2}s_{1,j}P(j,K,\sigma)+v_{1}P(K-1,K,\sigma) & \sum_{j=1}^{K-2}s_{1,j}2^{K-1}P(K,K,\sigma)+v_{1}2^{K-1}P(K,K,\sigma)\\
\vdots & \vdots\\	
\sum_{j=1}^{K-2}s_{K-2,j}P(j,K,\sigma)+v_{K-2}P(K-1,K,\sigma) & \sum_{j=1}^{K-2}s_{K-2,j}2^{K-1}P(K,K,\sigma)+v_{K-2}2^{K-1}P(K,K,\sigma)\\
\hline\\
\sum_{j=1}^{K-2}u_{j}P(j,K,\sigma)+\delta^{-1}P(K-1,K,\sigma) & 
\sum_{j=1}^{K-2}u_{j}2^{K-1}P(K,K,\sigma)+\delta^{-1}2^{K-1}P(K,K,\sigma)
\end{array}
\right ), \\\label{B}
\end{eqnarray}
where $s_{ij}$ are the entries of the matrix $\mathbf{S}$, and $v_j$ are the elements of the vector $\mathbf{v}$ defined in (\ref{N}).
The entry $E_{ij}$ of the matrix $\mathbf{E}$ is just the probability to jump from a node at distance $i$ from $n$, to a node at distance $j$ from $n$. Since we took as starting point of the stochastic process the node $n$, that corresponds to be in the state $0$ (i.e., at distance $0$), we consider only the first row of $\mathbf{E}$, that represents the probability to jump from $n$ toward a state at distance one (first column), or the opposite branch (second column).

\section{Mean first-passage times on $\mathcal{G}^{(p)}$}\label{sec:a}
We start our discussion referring to the case $p=2$, and then we are going to generalize the results on $p>2$.
The finite difference equation (\ref{eq:rec}),  can be solved analytically obtaining the following expression for the mean time $\tau(d)$ to first arrive to a given node at distance $d$:
\begin{eqnarray}
\tau(d)&=&\tau(1)+\tau(1)\left[\frac{x^{K-1}(x-1)}{2(x^{K-1}-1)}\right] \sum _{j=1}^{d-1} \prod _{l=1}^{j-1} A(j,K) \\
&=&\tau(1)+\tau(1)\left[\frac{x^{K-1}(x-1)}{2(x^{K-1}-1)}\right]\sum_{j=1}^{d-1} \frac{ 2^{1-j} x^j \left(x^K-1\right) \left(x^K-x\right)}{\left(x^j-x^K\right) \left(x^j-x^{K+1}\right)}\nonumber\\
&=& \tau(1)+\tau(1) \left[ \frac{x^K-x^{K-1}} {x^{K-1}-1} \right] (x^K-1)(x^K-x)\sum_{j=1}^{d-1}\frac{2^{-j}x^j}{\left(x^j-x^K\right) \left(x^j-x^{K+1}\right)}\nonumber\\
&=&\tau(1)\left[1+x^K(x-1)(x^K-1) \sum_{j=1}^{d-1}\frac{x^j}{2^{j}\left(x^j-x^K\right) \left(x^j-x^{K+1}\right)} \right],\label{td}
\end{eqnarray}
with $\tau(1)$ the mean time to first reach a node at distance $1$, on a graph with $2^K$ nodes.


In particular, let us focus on the case $d=K$, then  Eq.~\ref{td} becomes
\beas
\tau(K)&=&\tau(1) \left[1+x^K(x-1)(x^K-1) \right]\sum_{j=1}^{K-1}\frac{2^{-j}x^j}{\left(x^j-x^K\right) \left(x^j-x^{K+1}\right)},\nonumber
\eeas
that is
\be\label{ratiok1}
\frac{\tau(K)}{\tau(1)}= 1+[x^K(x-1)(x^K-1)]\sum_{j=1}^{K-1}\frac{2^{-j}x^j}{\left(x^j-x^K\right) \left(x^j-x^{K+1}\right)},
\ee
where, we recall, $x=2^{2\sigma-1}$. 

Our goal is to study the behavior of the ratio $\tau(K)/ \tau(1)$ with respect to $\sigma$ and to $K$. First of all we can rewrite the Eq.~\ref{ratiok1} in a more convenient way, that is
\begin{eqnarray}
\frac{\tau(K)}{\tau(1)}&=& 1+(x-1)\frac{x^K-1}{2^K}\sum_{m=1}^{K-1}\frac{(2x)^m}{(1-x^m)(1-x^{m+1})}\nonumber\\
&=& 1+(x-1)\frac{x^K-1}{2^K}\Big(\sum_{m=1}^{\overline{m}}\frac{(2x)^m}{(1-x^m)(1-x^{m+1})}+\sum_{m=\overline{m}}^{K-1}\frac{(2x)^m}{(1-x^m)(1-x^{m+1})}\Big),\label{b}
\end{eqnarray}
where $\overline{m}<K$ is a suitable integer number.
Since we are interested in the asymptotic behavior of $\tau(K)/\tau(1)$, we can fix $K\gg 1$ and $\overline{m}\gg 1$, the latter being independent of $K$. Therefore, we can approximate the second sum in Eq.~\ref{b} as
\begin{equation}
	\sum_{m=\overline{m}}^{K-1}\frac{(2x)^m}{(1-x^m)(1-x^{m+1})}\approx	\sum_{m=\overline{m}}^{K-1}\frac{(2x)^m}{x^{2m+1}}=\frac{1}{x}\sum_{m=\overline{m}}^{K-1}2^mx^{-m}=\frac{1}{x-2}\Big(\frac{2^{\overline{m}}}{x^{\overline{m}}}-\frac{1}{2^{K(2\sigma-2)}}\Big). \label{sommatoriaHFM}
\end{equation}
Plugging Eq.~\ref{sommatoriaHFM} into Eq.~\ref{ratiok1}, we can state that, for large $K$ the ratio $\tau(K)/\tau(1)$ can be approximated as
\be
\frac{\tau(K)}{\tau(1)}\approx 1+\frac{x-1}{x-2}\frac{x^K-1}{2^K}\Big(C(\overline{m},x)+\frac{2^{\overline{m}}}{x^{\overline{m}}}-\frac{1}{2^{K(2\sigma-2)}}\Big),\label{approx}
\ee
where $C(\overline{m},x)$ is the contribution of the first term of the right-hand side of Eq.~\ref{b}.

We have now to focus on two separated cases corresponding to $\sigma\in(1/2,1)$ and to $\sigma=1$, respectively. In the former case, letting $K$ grow, and considering the infinite size limit, we show the existence of an asymptotic value for $\tau(K)/\tau(1)$ as
\be
\frac{\tau(K)}{\tau(1)}\xrightarrow[K \rightarrow \infty]{}1-\frac{x-1}{x-2}=\frac{1}{2-x}, \qquad x\in(1,2).\label{asimp}
\ee
In the latter case $\sigma=1$ (that is $x=2$), and the sum in Eq.~\ref{sommatoriaHFM} grows linearly with $K$ in such a way that Eq.~\ref{approx} becomes
\be \label{analitico1}
\frac{\tau(K)}{\tau(1)}\approx 1+\frac{1}{2}K.
\ee

\begin{figure}[h!]
	\includegraphics[width=12cm]{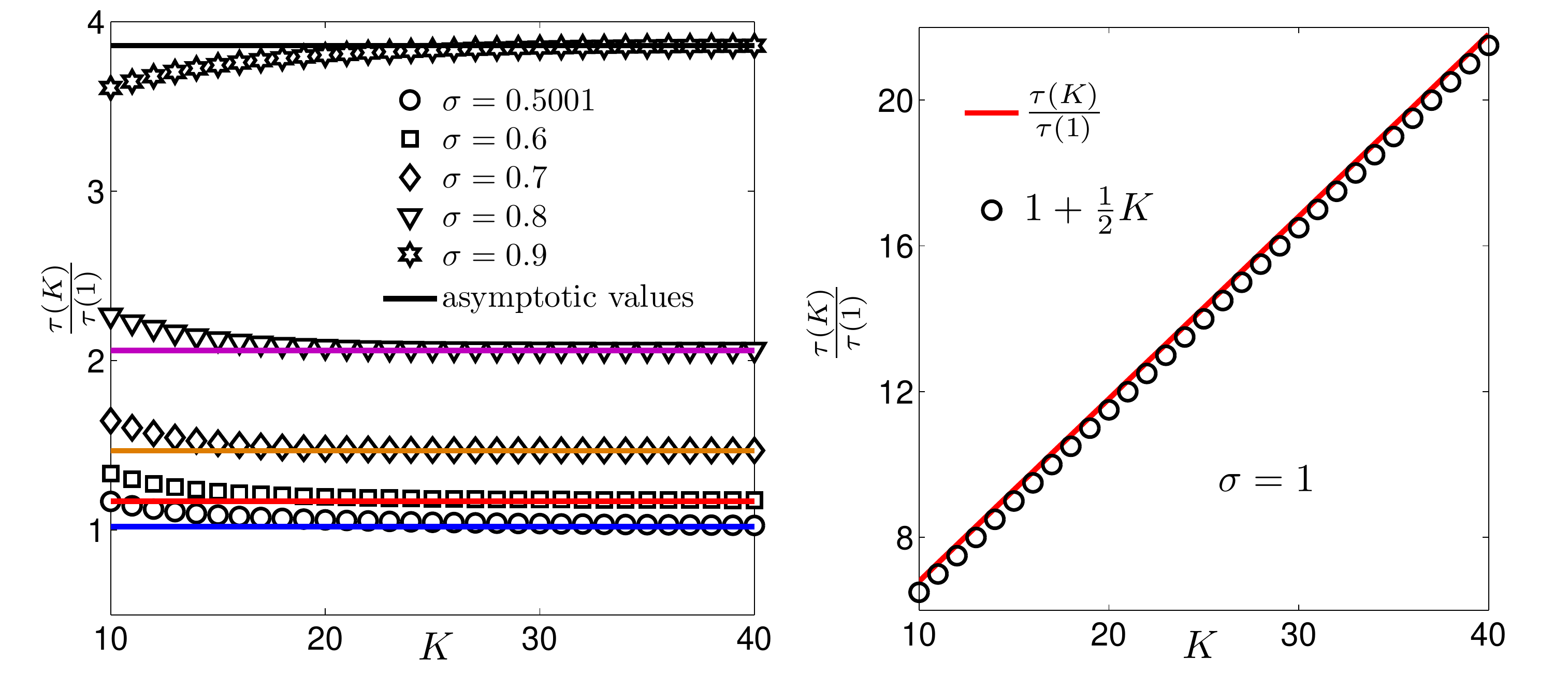}		
	\caption{(Color
		online) Left panel: the ratio $\tau(K)/\tau(1)$ is plotted versus $K\in[10,40]$ for several choices of $\sigma\in(1/2,1)$, as shown in the legend. Data obtained by numerical evaluating Eq.~\ref{td} (symbols) are compared with the asymptotic value obtained analytically in Eq.~\ref{asimp}. Right panel: the ratio $\tau(K)/\tau(1)$ is plotted versus $K\in[10,40]$ for $\sigma=1$. As one can see, its behavior is qualitatively different from the one shown in the left panel and characterizing the case $\sigma<1$, in fact, here $\tau(K)/\tau(1)$ grows linearly with $K$. Data obtained by numerical evaluating Eq.~\ref{td} ($\circ$) are compared with the result obtained analytically in Eq.~\ref{analitico1}.  }\label{fig:as}
\end{figure}

It is worth noticing that as $\sigma\rightarrow 1/2$, the asymptotic value of the ratio tends to $1$, and this confirms that as $\sigma$ tends to its lower bound, and the size of the graph grows, the graph $\mathcal{G}^{(2)}$ recovers a fully connected network with homogeneous weights of the links. Conversely, when $\sigma=1$, the asymptote disappears and the linear growth of the ratio highlights that the dependence of $\tau(1)$ and of $\tau(K)$ on the system size is qualitative different. 

\begin{figure}[h!]
 	\includegraphics[width=16cm]{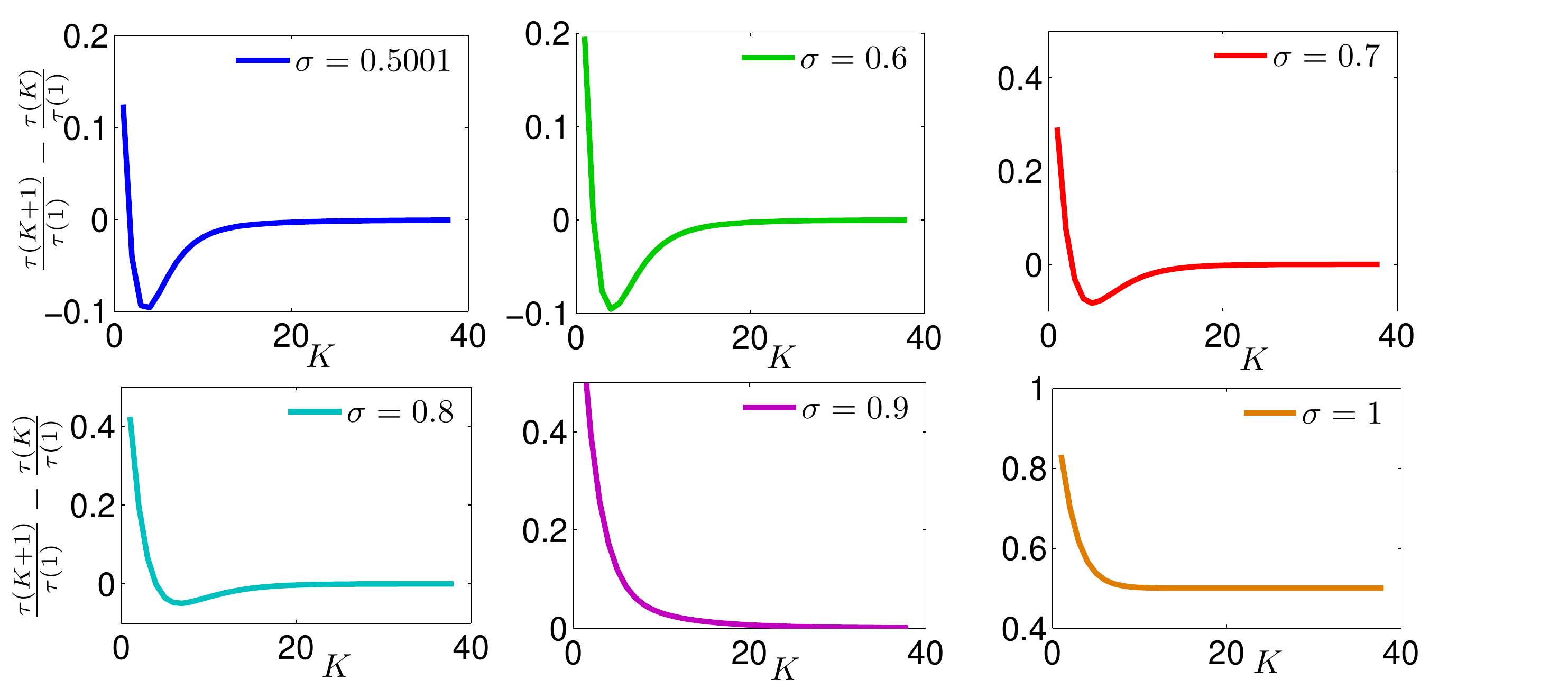}		
 	\caption{(Color
 		online) Trend of the difference $\tau(K+1)/\tau(1)-\tau(K)/\tau(1)$ as the size increases ($K\in[2,40]$ for different values of $\sigma\in(0.5,1]$. As one can see, the behavior of the trend changes with respect t $\sigma$, and it outlines the presence of  a critical size such that the variation of the ratio of the times }\label{fig:diff}
 \end{figure}

Finally, we deepen another feature characterizing the dependence of $\tau(K)/\tau(1)$ from $K$: as one can see in Fig.~\ref{fig:tKt1}, $\tau(K)/\tau(1)$ exhibits a peak, in such a way that for relatively small values of $K$ the ratio between times grows with $K$, while for relatively large values it decreases asymptotically towards $1/(2-x)$.
To this aim, let us focus on the difference $\Delta(K) = \tau(K+1)/\tau(1)-\tau(K)/\tau(1)$ and let us study its behavior as a function of $K$ and of $\sigma$. Basically, $\Delta(K)$ represents the numerical derivative of the ratio and the peak of $\tau(K)/\tau(1)$ corresponds to the vanishing of $\Delta(K)$.
As one can see in Fig.~\ref{fig:diff}, $\Delta(K)$ does change sign and this effect is  especially pronounced for small values of $\sigma$. On the other hand, when $\sigma$ gets closer to $1$, $\Delta(K)$ vanishes at larger and larger values of $K$ and its minimum value is, in magnitude, smaller and smaller. When $\sigma=1$, $\Delta(K)$ never vanishes and tends to $1/2$ asymptotically.
This is perfectly consistent with the results outlined above.
This discussion can be extended to the case $p>2$; in fact, recalling that $x=p^{p-3+2\sigma}$, and starting from
\begin{equation}
\frac{\tau(K)}{\tau(1)}=1+(x-1)\frac{x^K-1}{p^K}\sum_{m=1}^{K-1}\frac{(px)^m}{(1-x^m)(1-x^{m+1})},
\end{equation}
we can again study the trend of the sum in such a way that
\begin{eqnarray}
\frac{\tau(K)}{\tau(1)} &\approx & 1+ (x-1)\frac{x^K-1}{p^K}\left[C(x,\overline{m})+\sum_{m=\overline{m}}^{K-1}\frac{(px)^m}{(1-x^m)(1-x^{m+1})}\right]\nonumber\\
&\approx & 1+ (x-1)\frac{x^K-1}{p^K}\left[C(x,\overline{m})+\sum_{m=\overline{m}}^{K-1}\frac{(px)^m}{x^{2m+1}}\right]\nonumber\\
&=& 1+\frac{x-1}{x-p}\frac{x^K-1}{p^K}\Big[C(x,\overline{m})+\left(\frac{p}{x}\right)^{\overline{m}}-\left(\frac{p}{x}\Big)^K\right],\label{pas}
\end{eqnarray}
 where $\overline{m}\gg 1$,  $\overline{m}\leq K$ and $K\gg 1$.
Now, considering the limit $k\rightarrow\infty$ of Eq.~\ref{pas}, we obtain
 \be
\frac{\tau(K)}{\tau(1)}\approx p^{K(p-4+2\sigma)} \xrightarrow[K\rightarrow\infty]{}\infty,
 \ee
 since, in this case $x>p$, for every choice of $\sigma\in(1/2,1]$ and $p>2$.
 \newline
 These results corroborate the estimates presented in Sec.~\ref{sec:hittingp} about the trend of the ratio of the mean first-passage times for graphs with $p>2$.

\section{Splitting probabilities on $\mathcal{G}^{(p)}$} \label{sec:c}
In this appendix we summarise the results found for the splitting probabilities on $\mathcal{G}^{(p)}$, in analogy with the treatment followed for $\mathcal{G}^{(2)}$ and presented in Sec.~\ref{sec:splitting}.
Once again, fixed a generic site $i$, we are going to define a Markov chain whose states are $\{0,1,...,K\}$, that is all the possible distances from $i$. In particular, as already shown for $\mathcal{G}^{(2)}$ we are going to  consider the states $1$ and $K$ as absorbing states, while, all the others are transient. In this way we build up a transition matrix  $\mathbf{M}$ whose entry $M_{lj}$ is the probability to jump from a node at distance $l$ from $i$ toward a node at distance $j$ from $i$. We can now proceed using the same arguments proposed in Sec.~\ref{sec:splitting}, since the only difference between this model and the one with $p=2$ is in the specific probabilities necessary to jump from a state to another. In this case, as one expects, we recover a completely different result: the probability $P(K|1)$ of being absorbed at distance $K$, where $K$ is the maximum reachable distance is, for all $p>2$ always close to $0$, while $P(1|K)$ is always almost or equal to $1$. This means that a stochastic process starting its path on a generic node $n$ will almost surely be absorbed at sites at distance $1$, while the probability of being absorbed at any node at distance $K$ is vanishing, although their number grows exponentially with $K$ as $N=p^K$. This confirms again that in $\mathcal{G}^{(p)}$ the overall weight of the longest links is negligible with respect to the the overall weight of the shortest links.

\begin{figure}[h!]
	\includegraphics[width=10cm]{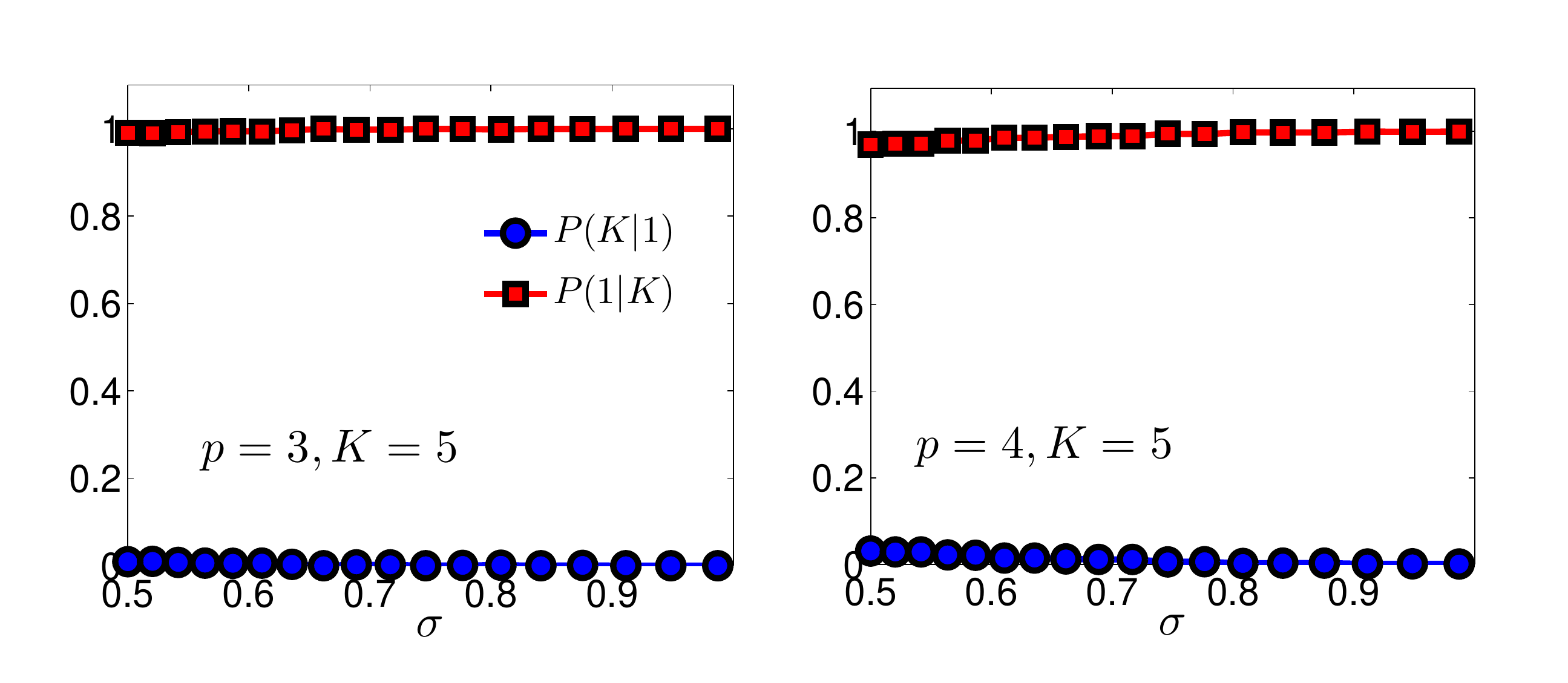}
	\caption{(Color
		online) Splitting probabilities $P(1|K)$ and $P(K|1)$ for $p=3$ (left panel) and $p=4$ (right panel), as a function of $\sigma$ and with $K=5$ fixed. Data points are obtained via numerical simulations with average over $10^3$ realizations. Notice that, for all values of $\sigma$, the stochastic process never reached any state at distance $K$ before the state at distance $1$. This is markedly different than the behaviour obtained for $p=2$ and shown in Fig.~\ref{fig:criAsympt}.}		
\end{figure}

\end{document}